\documentclass[runningheads]{llncs}
\usepackage{amsfonts}
\usepackage{amsmath}
\usepackage{graphicx}
\usepackage{mytodo}
\usepackage{tabularx}
\usepackage{footnote}
\makesavenoteenv{tabularx}
\PassOptionsToPackage{hyphens}{url}\usepackage[hidelinks]{hyperref}
\usepackage{multicol}
\usepackage{hyperref}

\begin{document}
\makeatletter
\renewcommand*\l@author[2]{}
\renewcommand*\l@title[2]{}
\makeatletter
\title{Exploring the Dynamics of Data Transmission in 5G Networks: A Conceptual Analysis}

\author{Nikita Smirnov\inst{1} \and Sven Tomforde\inst{2}}
\authorrunning{Nikita Smirnov, Sven Tomforde}
\titlerunning{Exploring the Dynamics of Data Transmission in 5G Networks}

\institute{Kiel University, Christian-Albrechts-Platz 4, 24118 Kiel, Germany\inst{1,}\inst{2} \\
\email{nsm@informatik.uni-kiel.de, st@informatik.uni-kiel.de}\\
\url{https://www.ins.informatik.uni-kiel.de/}}

\maketitle

\pagenumbering{gobble}
\begin{abstract}\normalsize
    This conceptual analysis examines the dynamics of data transmission in 5G networks. It addresses various aspects of sending data from cameras and LiDARs installed on a remote-controlled ferry to a land-based control center. The range of topics includes all stages of video and LiDAR data processing from acquisition and encoding to final decoding, all aspects of their transmission and reception via the WebRTC protocol, and all possible types of network problems such as handovers or congestion that could affect the quality of experience for end-users. A series of experiments were conducted to evaluate the key aspects of the data transmission. These include simulation-based reproducible runs and real-world experiments conducted using open-source solutions we developed: "Gymir5G" - an OMNeT++-based 5G simulation and "GstWebRTCApp" - a GStreamer-based application for adaptive control of media streams over the WebRTC protocol. One of the goals of this study is to formulate the bandwidth and latency requirements for reliable real-time communication and to estimate their approximate values. This goal was achieved through simulation-based experiments involving docking maneuvers in the Bay of Kiel, Germany. The final latency for the entire data processing pipeline was also estimated during the real tests. In addition, a series of simulation-based experiments showed the impact of key WebRTC features and demonstrated the effectiveness of the WebRTC protocol, while the conducted video codec comparison showed that the hardware-accelerated H.264 codec is the best. Finally, the research addresses the topic of adaptive communication, where the traditional congestion avoidance and deep reinforcement learning approaches were analyzed. The comparison in a sandbox scenario shows that the AI-based solution outperforms the WebRTC baseline GCC algorithm in terms of data rates, latency, and packet loss.
    \vspace{1mm}
    
\end{abstract}

\clearpage
\begin{center}
  \textbf{Acknowledgments}
\end{center}

This research has been funded by the German Federal Ministry for Digital Affairs and Transport (Bundesministerium für Digitales und Verkehr) within the project "CAPTN Förde 5G", funding guideline: "5G Umsetzungsförderung im Rahmen des 5G-Innovationsprogramms", funding code: 45FGU139\_H. The authors acknowledge the financial support of the BMDV.

\clearpage
\tableofcontents

\clearpage
\addtocontents{toc}{\vspace{\normalbaselineskip}}
\pagenumbering{arabic}
\section{Introduction}\label{sec1_introduction}
The rapid development of 5G technology has increased the potential reliability of remote-controlled units (RCUs). Real-time communication (RTC) in 5G-based RCUs is critical to ensure seamless communication between the operator and the unit, which requires low latency and high reliability. As more data is often generated than network capacity is available, the system needs to select and adapt the communicated data at runtime based on changing conditions. This data serves as the basis of situational awareness for human operators at a central control center and contributes to the agent's decision-making capabilities, facilitating the transition from remote control to autonomy.

The main source of situation awareness is multiple video and point cloud streams captured by sensors installed on the unit. They are located at the different corners of the unit and together provide a 360-degree view. These streams operate continuously, placing substantial demands on the network infrastructure, requiring a high throughput rate and sub-second latency.

\begin{figure}[!b]
\center
\includegraphics[width=1.0\columnwidth, height = 6cm]{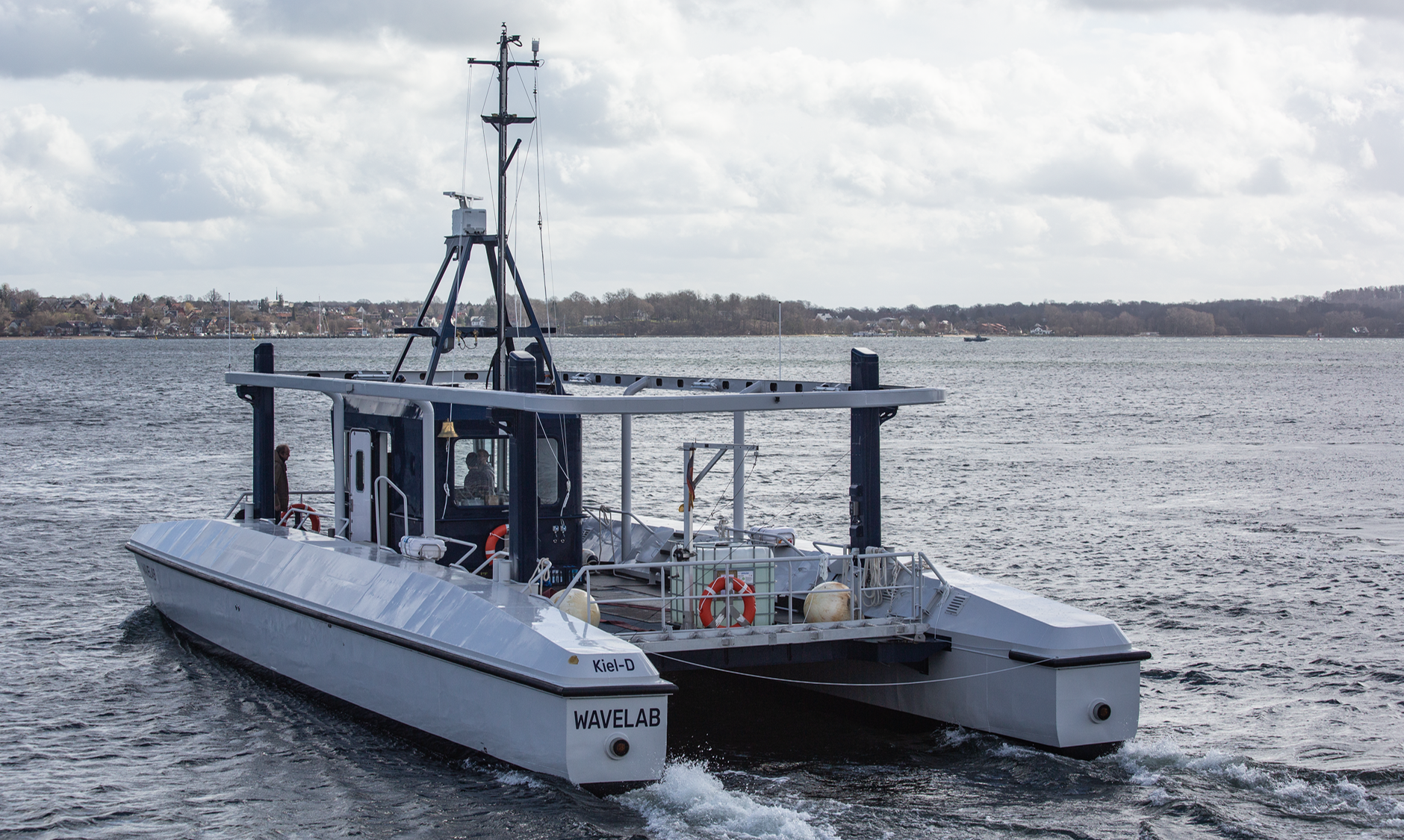}
\caption{5G remote-controlled test ferry ”Wavelab”.} 
\label{figure:wavelab}
\end{figure}

The unmanned ferry "Wavelab" (see Fig.~\ref{figure:wavelab}) serves as an example of such an RCU in the research project CAPTN Fjord5G~\cite{foerde5g}. Equipped with 5G routers, it streams data to the remote control center on the shore while sailing in the Bay of Kiel and the Kiel Canal. This is especially important as the navigation system learns to perform complicated maneuvers such as docking and requires reliable sensor data transmission. The "Wavelab" ferry serves as a test platform for a larger unmanned ferry designed for passenger transport in the Bay of Kiel.

This concept study deals with various topics considering different aspects of data transmission while also aiming to investigate the limits and peculiarities of 5G networks in the scope of the Fjord5G project. Namely, this study tries to answer the following questions:
\begin{itemize}
\item How much data could be transmitted via 5G in general and in the Kiel water area? 
    \item What network scenarios could occur?
    \item How exactly should the live sensor data be transmitted in real-time to provide the best quality of experience in the control center?
    \item How should the live sensor data be preprocessed to meet the quality/latency trade-off?
    \item How could real-time live data transmission be adapted with the help of AI?
\end{itemize}

\subsection{Simulation platform}
\begin{figure}[!hb]
  \centering
    \includegraphics[width=1.0\textwidth, height=8cm]{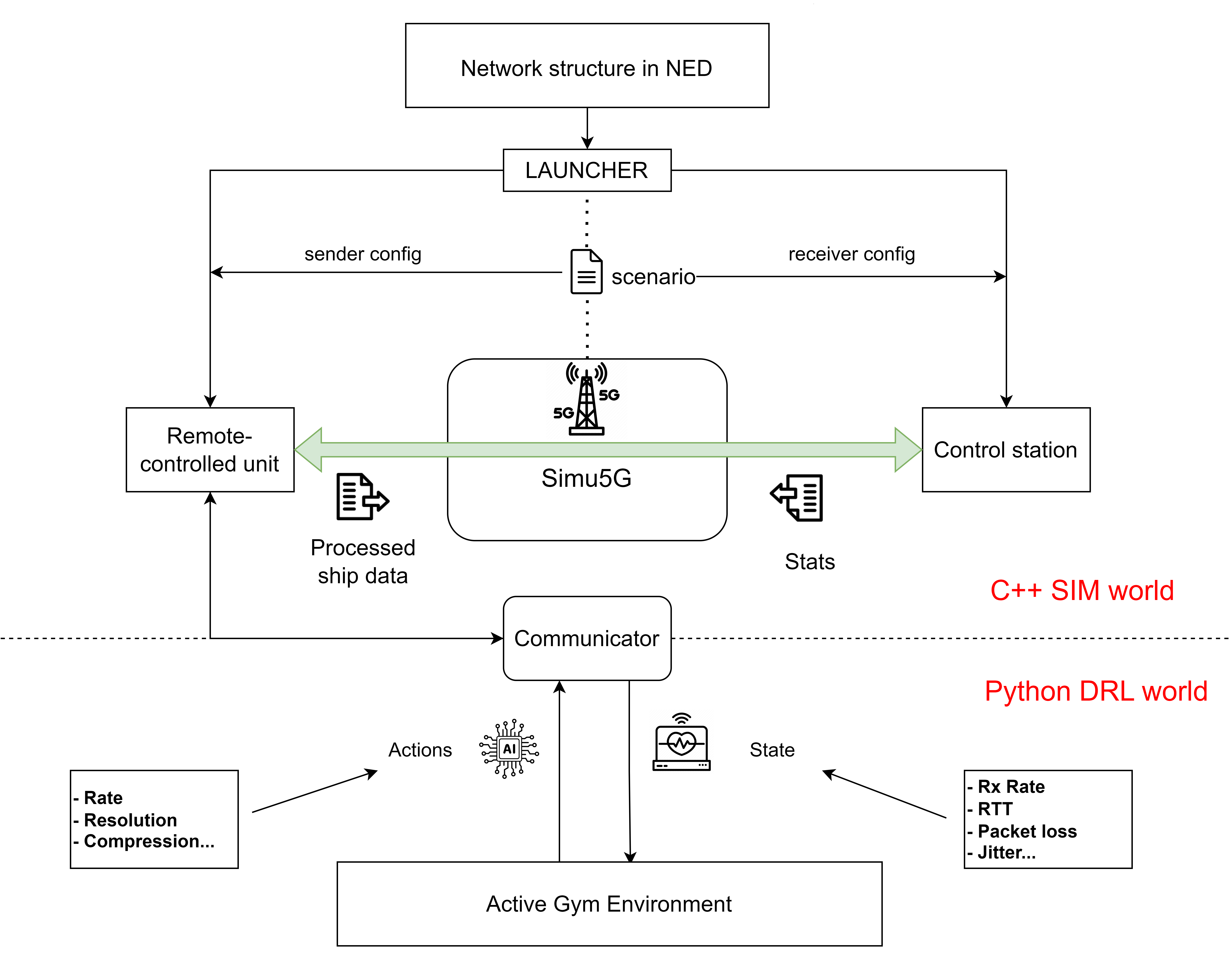}
    \caption{Gymir5G workflow.}
    \label{figure:gymir5g}
\end{figure}

All simulation-based experiments for this study were performed in the "Gymir5G" platform, which we developed as part of the CAPTN Fjord5G project\footnote{Available at \url{https://github.com/gehirndienst/gymir5g}, last accessed: \today}. It stands out as a software tailored to the study of the optimization of media data transmission in wireless networks, with a particular focus on the use of deep learning technologies to specifically bridge the gap between simulated scenarios and real-world applications. The workflow is presented in Fig.~\ref{figure:gymir5g}

At its core, Gymir5G is written in C++ and based on the OMNeT++ discrete event simulator ecosystem~\cite{omnetpp}, using the capabilities of the additional libraries developed on top of OMNeT++ to simulate different wireless network characteristics. This choice provides a powerful, flexible framework for modeling and analyzing complex communications systems. GymirDRL, a Python-based subproject of Gymir5G, introduces a Deep Reinforcement Learning (DRL) framework. In the world of technology, DRL is an example of a Deep Learning (DL) technology that aims to learn from experience, paving the way for smarter decision-making in dynamic environments. GymirDRL allows for active online data optimization within the simulation and thus effectively trains such active models.

A special highlight of Gymir5G is its emphasis on real-time communication for the web (WebRTC) protocol~\cite{webrtc_def}. WebRTC, the backbone of real-time multimedia communication on the Web, is implemented in Gymir5G with its basic key features such as congestion control (GCC algorithm and TWCC feedback packets), retransmission (NACK), and redundancy mechanisms (FEC). This incorporation enhances the platform's ability to reflect real-world communication scenarios more accurately.

\subsection{Real testbed}\label{section:1.2_real_testbed}
Real experiments were conducted on the development environment of the AhoyRTC Director platform developed by AhoyRTC GmbH~\cite{ahoy}. It is a platform for playback of WebRTC feeds attached by viewers. It is actively used in production to transmit video streams from cameras installed on the Wavelab ferry. 

The streams are streamed using the Python application "Gstwebrtcapp", also developed by our working group within the CAPTN Fjord5G project\footnote{Available at \url{https://github.com/gehirndienst/gstwebrtcapp}, last accessed: \today}. It is used to stream video from the given source via the GStreamer pipeline~\cite{gstreamer} to the WebRTC client with the possibility to control the video quality on the fly. This includes bitrate, resolution, framerate, and other encoder settings. The WebRTC statistic is obtained via internal GStreamer functions and grabbed directly from the viewer.

\subsection{Text formatting}
In our text, we employ specific coloring for itemized lists to denote different types of analysis: we use green checkboxes \textcolor{green}{\checkmark} to represent our conclusions, blue bullets \textcolor{blue}{\textbullet} for technical discussions, and red minus signs \textcolor{red}{$-$} to highlight problems encountered.

\clearpage
\section{Data transmission in 5G networks}
\subsection{Assessing the quality of data transmission}
We assess the quality of data transmission in general by looking at three main metrics: a) goodput, b) latency, and c) packet loss.

Speed, or network throughput/goodput, is a critical metric for evaluating the efficiency of data transmission and represents the actual useful data received at the application layer over the network in a given period. It is measured in bits (kilobits, megabits) per second. A high throughput indicates a more efficient use of network resources, resulting in better quality of transmitted data.

Latency is another important metric for evaluating the quality of data transmission. Latency refers to the time delay between the initiation of a data transfer and the actual receipt of the data at its destination. It is commonly measured as round-trip time (RTT), which is the total time it takes for a packet of data to travel from the source to the destination and back. Low latency is particularly desirable in real-time applications such as video streaming, as it ensures minimal delays and a more responsive user experience. High latency results in noticeable delays and causes the video to stutter.

Packet loss rate (PLR) is the third key metric. Packet loss occurs when data packets traveling through the network do not reach their destination, and the receiver waiting for these packets cannot continue to properly decode or process the data. Packet loss can be caused by several factors, including network congestion, hardware failure, or problems with the underlying infrastructure. Excessive packet loss can result in degraded performance, retransmissions, and a poor user experience.

\subsection{Docking scenarios}
\begin{figure}[!t]
    \includegraphics[width=\textwidth, height = 6cm]{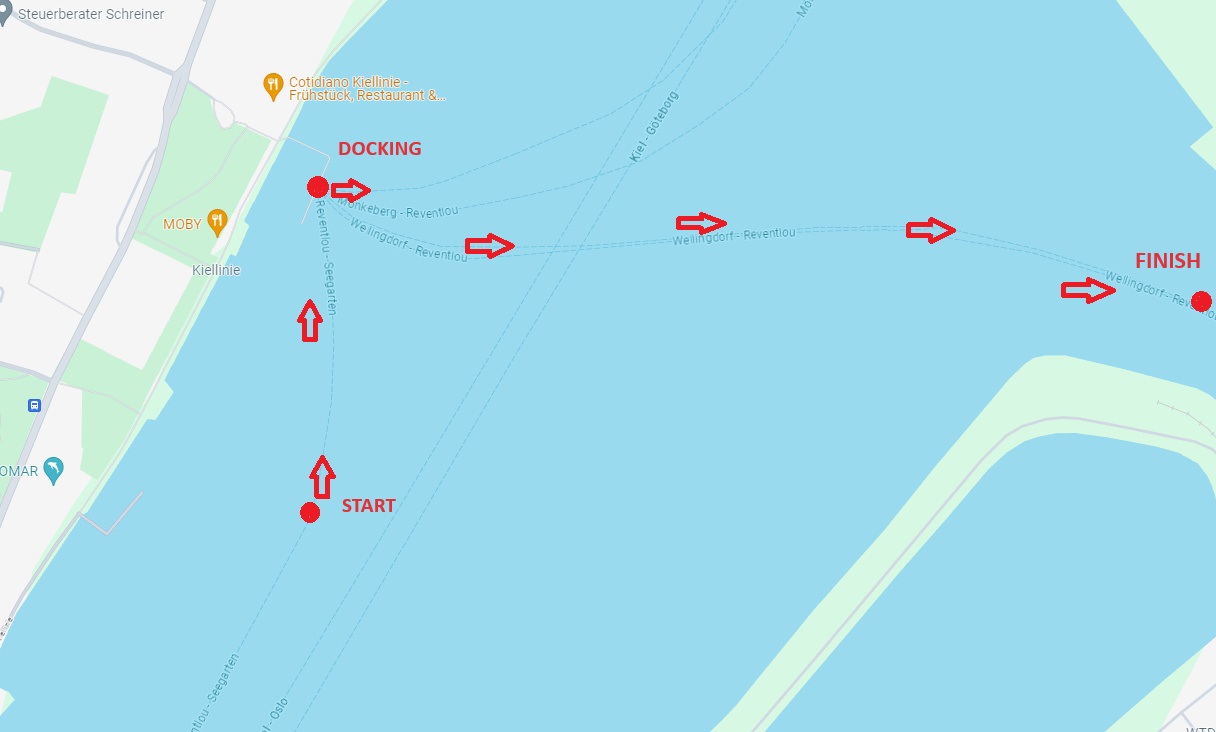}
    \caption{Reventlou docking scenario.}
    \label{figure:reventlou}
\end{figure}

\begin{figure}[!t]
    \includegraphics[width=\textwidth, height = 6cm]{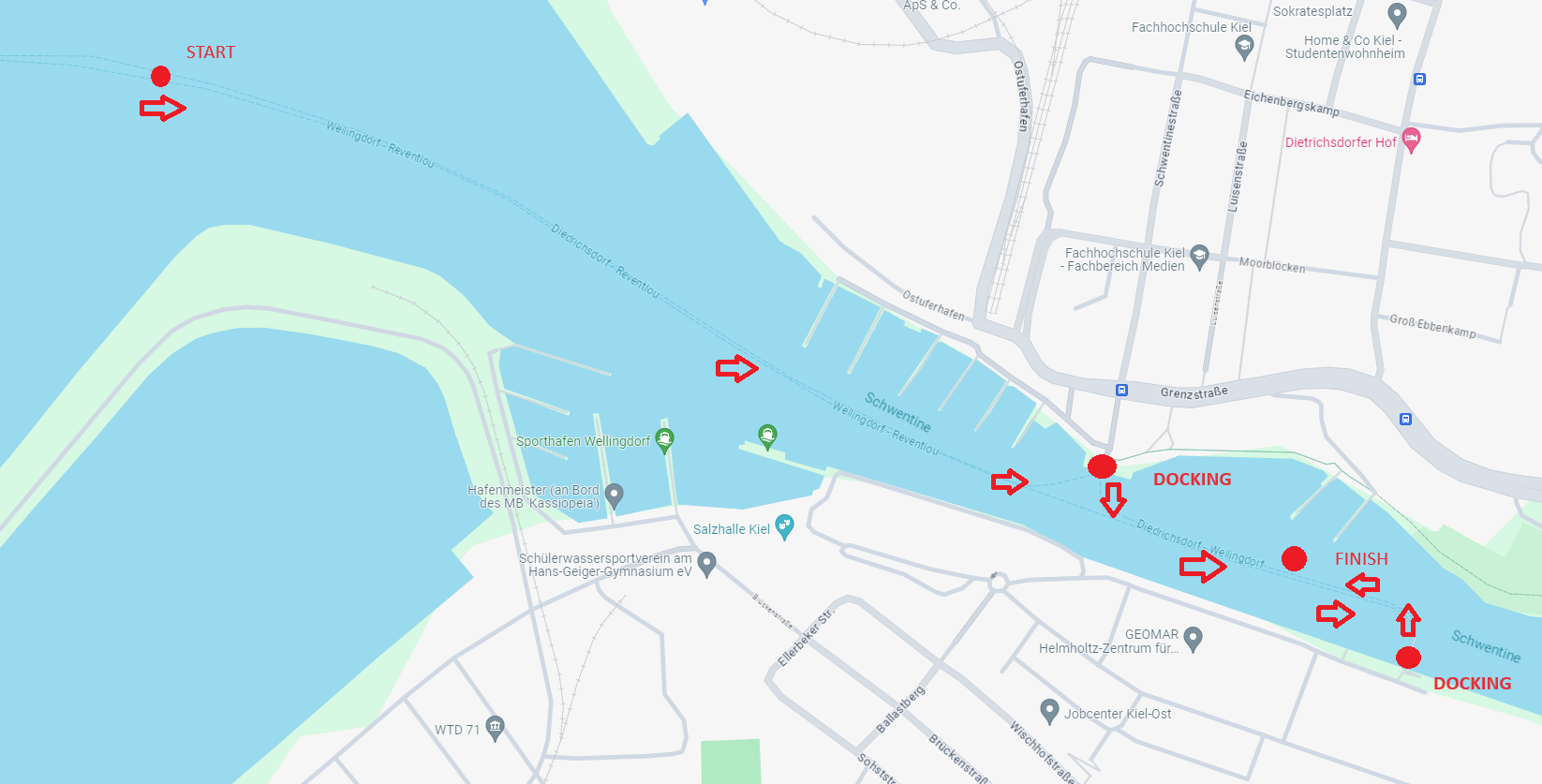}
    \caption{Dietrichsdorf and Wellingdorf docking scenarios.}
    \label{figure:ditrichsdorf}
\end{figure}

To estimate the aforementioned metrics for the Bay of Kiel, we have modeled two docking scenarios in Gymir5G for the "Schwentine" area. One of them is around the Reventlou on the left coast (see Fig.~\ref{figure:reventlou}) and another one is on the right coast with two docking stations Dietrichsdorf and Wellingdorf (see Fig.~\ref{figure:ditrichsdorf}). Those scenarios are referred to later in the text as REV and DIT.

We modeled the 5G network of the German operator Vodafone in the selected areas. We did not have direct access to the internal parameters of Vodafone's base stations or cell configurations, so we had to use an open-source portal called Cellmapper~\cite{cellmapper}. For the background traffic, we had to rely on other open-access datasets to model the network load. Two different use cases were studied: one with normal background traffic in both uplink and downlink directions, and one without background traffic to analyze the limits of 5G bandwidth.

\begin{figure}[!t]
\centering
    \includegraphics[width=0.65\textwidth, height = 7.5cm]{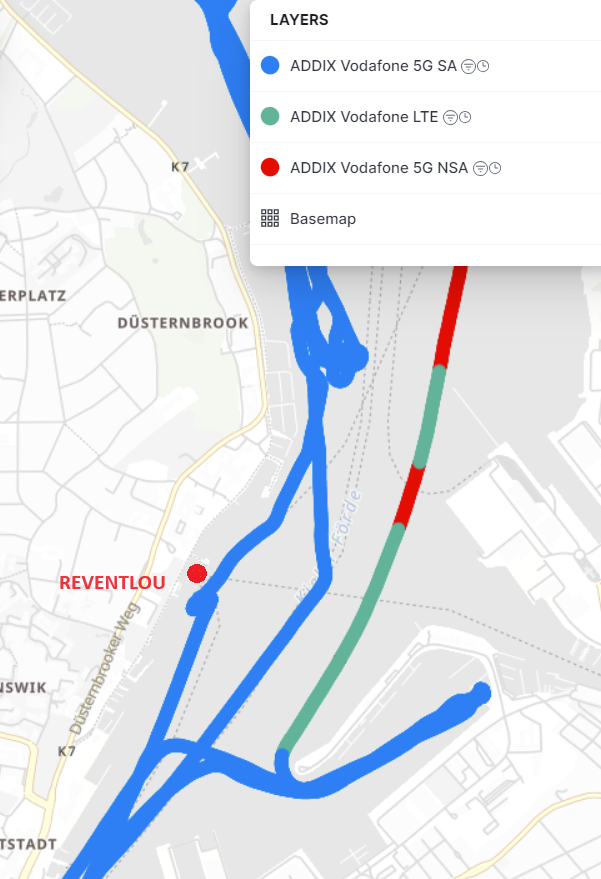}
    \caption{5G coverage by Vodafone operator in the "Schwentine" area of the Bay of Kiel.}
    \label{figure:5g_coverage}
\end{figure}

The REV scenario is characterized as a "good coverage" scenario. During docking, the 5G standalone, i.e., operate fully within 5G infrastructure (5G SA) connection is provided by 3 base stations located near the coast. The DIT scenario is characterized as a "bad coverage" scenario because the base stations are non-standalone (5G NSA), i.e., the core network runs under the 4G standard and they are far away from the coast. Additionally, there is the Kiel University of Applied Sciences, which creates a huge background traffic. Fig.~\ref{figure:5g_coverage} shows the 5G coverage of Vodafone in the "Schwentine" area, recorded from the real voyages of the Wavelab ferry.

For both scenarios, we also implemented a configuration with no background traffic to measure the limits of 5G networks. We transmitted 80 Mbits of data for this case, compared to 50 Mbits for the background case, because real-world measurements showed that 80-100 Mbits is the average 5G uplink bandwidth under good conditions. And 50 Mbits were chosen to test whether the 5G network could provide enough bandwidth to transmit the data from the \textbf{fast moving ship} for approximately 6-8 good-quality video streams and at least one LiDAR stream. For the REV, we also implemented a downlink test scenario where a LiDAR is installed on the Reventlou ship station and the ship receives this stream. The LiDAR streams are modeled on 2 Blickfiedl Cube 1 Outdoor sensors with a common average data rate of 20 Mbps (10 Mbps each). This experiment corresponds to the idea of an \textbf{intelligent pier}, where the data streamed by the sensors could be received by all passing ships. We have presented and compared all the results in the following subsection.

The last experiment deals with the multi-homing scenario for the REV. For this, we modeled new functionality in Gymir5G to create a device capable of using wireless and cellular communication stacks in parallel. We modeled access points around the Kiellinie near Reventlou and tried to divide the traffic into equal groups where one part flows over Wi-Fi and another part over the 5G communication link.

\begin{figure}[!t]
    \includegraphics[width=\textwidth, height = 6.3cm]{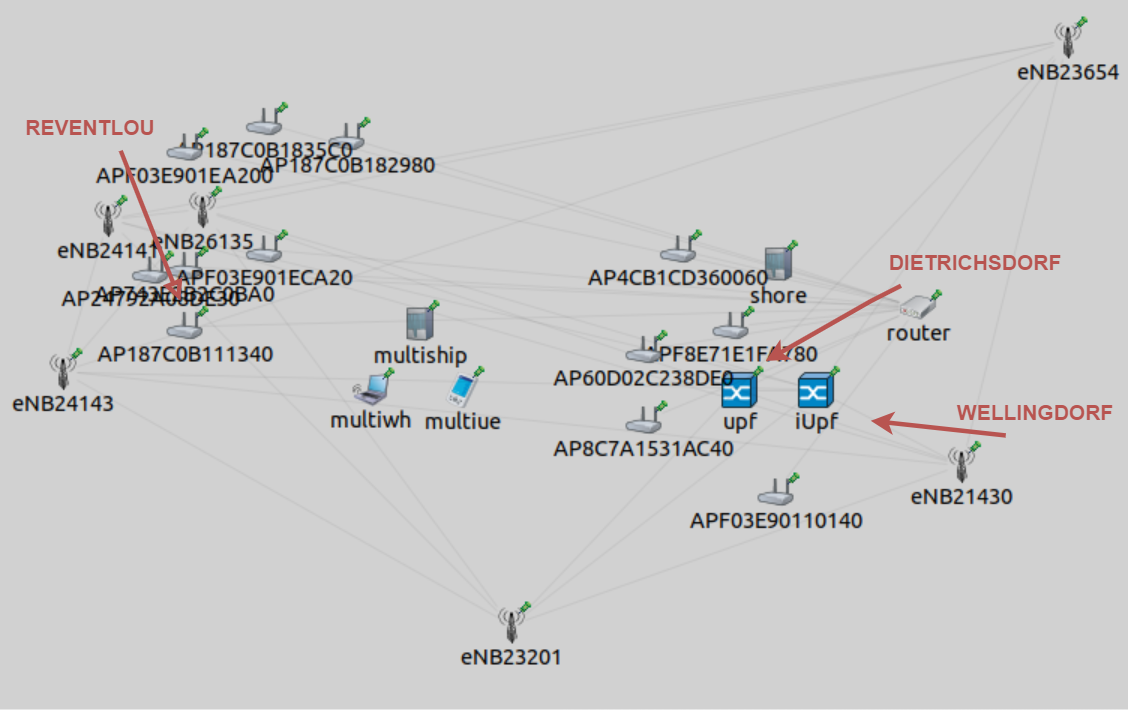}
    \caption{Gymir5G network scenario for the whole "Schwentine" area and all scenarios. eNB* stands for 5G SA (left) and NSA (right) stations. AP* means Wi-Fi access points. Corresponding docking stations are also marked in the picture.}
    \label{figure:sim_scen}
\end{figure}

The entire network for all scenarios is shown in Fig.~\ref{figure:sim_scen}. The REV 5G SA scenario is on the left, while the DIT 5G NSA scenario is on the right. It is already clear from the figure that the distance between the base stations and the moving agent is small in the case of REV and large in the case of DIT. The "multi*" constellation of devices represents the Wavelab with the wireless and cellular network stacks. The "shore" object represents the control center. By default, it is located at the Kiel University of Applied Sciences. The "Upf" objects represent a 5G User Plane Function (with an interface), an important middle layer that connects the radio and network 5G stacks. 

\subsection{Results}
\subsubsection{Uplink}
\begin{figure}[!t]
    \includegraphics[width=\textwidth, height = 6.8cm]{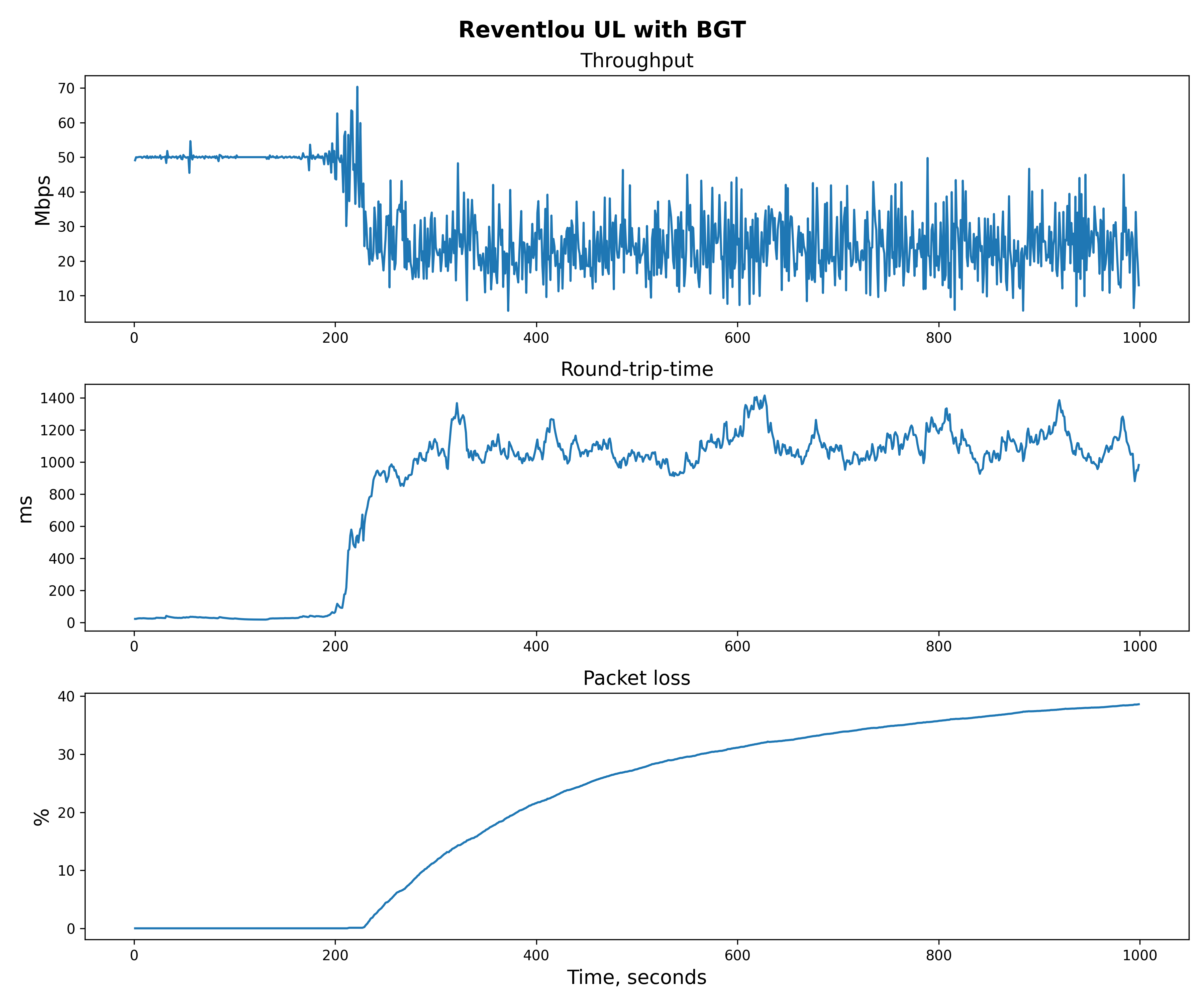}
    \caption{REV UL run with the background traffic.}
    \label{figure:rev_bg}
\end{figure}
\begin{figure}[!t]
    \includegraphics[width=\textwidth, height = 6.8cm]{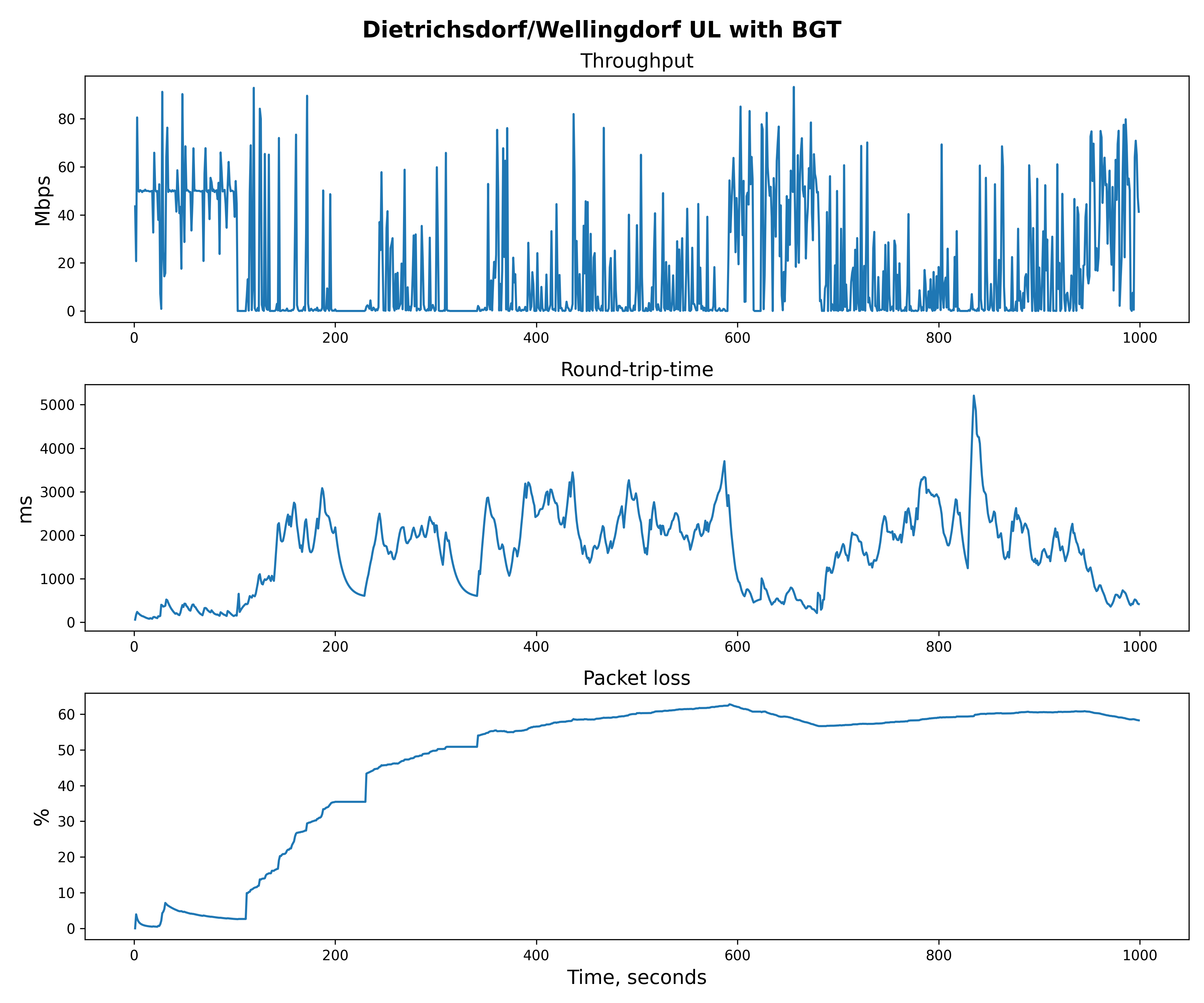}
    \caption{DIT UL run with the background traffic.}
    \label{figure:dit_bg}
\end{figure}
\begin{figure}[!ht]
    \includegraphics[width=\textwidth, height = 6.8cm]{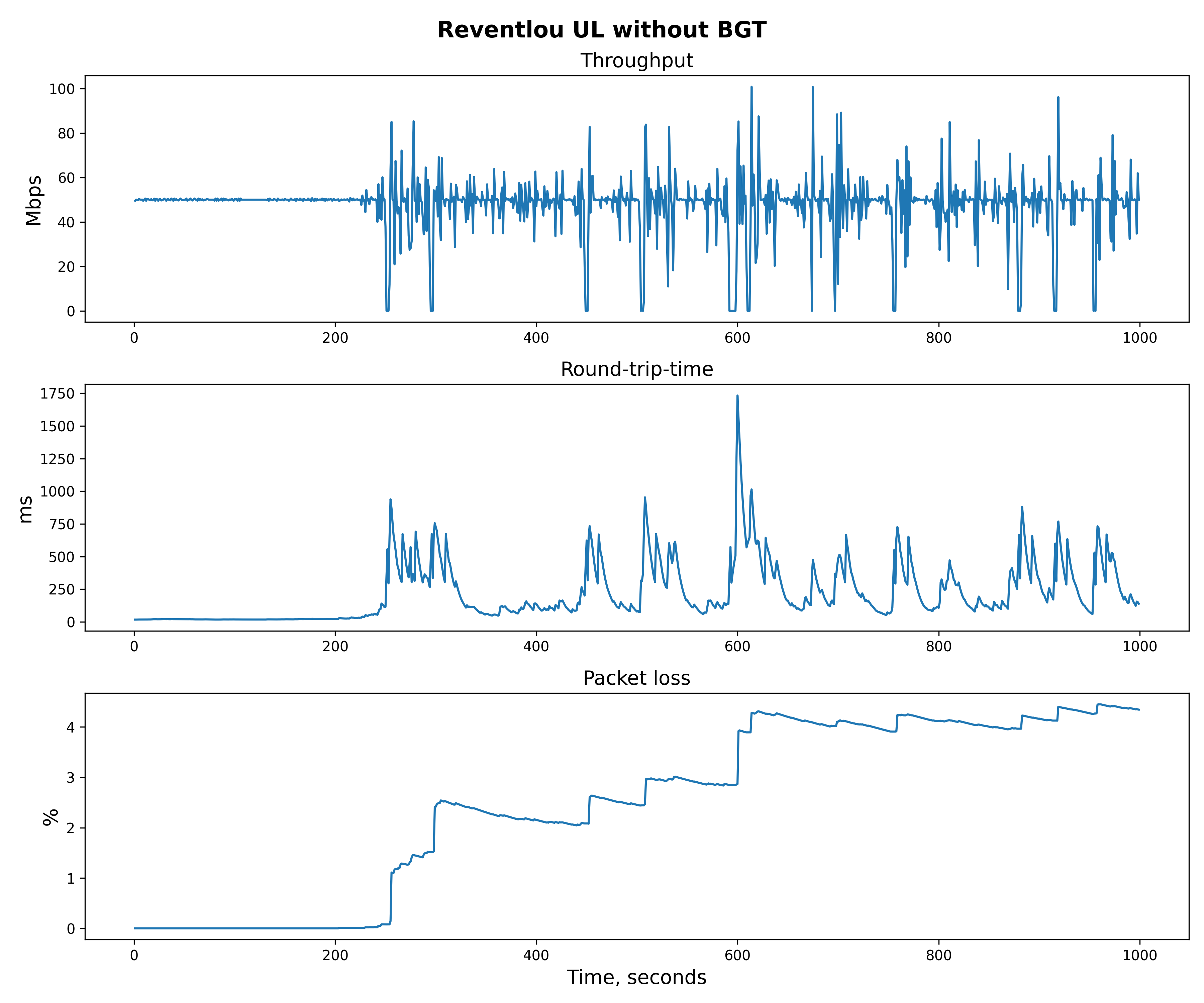}
    \caption{REV UL run without the background traffic.}
    \label{figure:rev_nt}
\end{figure}
\begin{figure}[!ht]
    \includegraphics[width=\textwidth, height = 6.8cm]{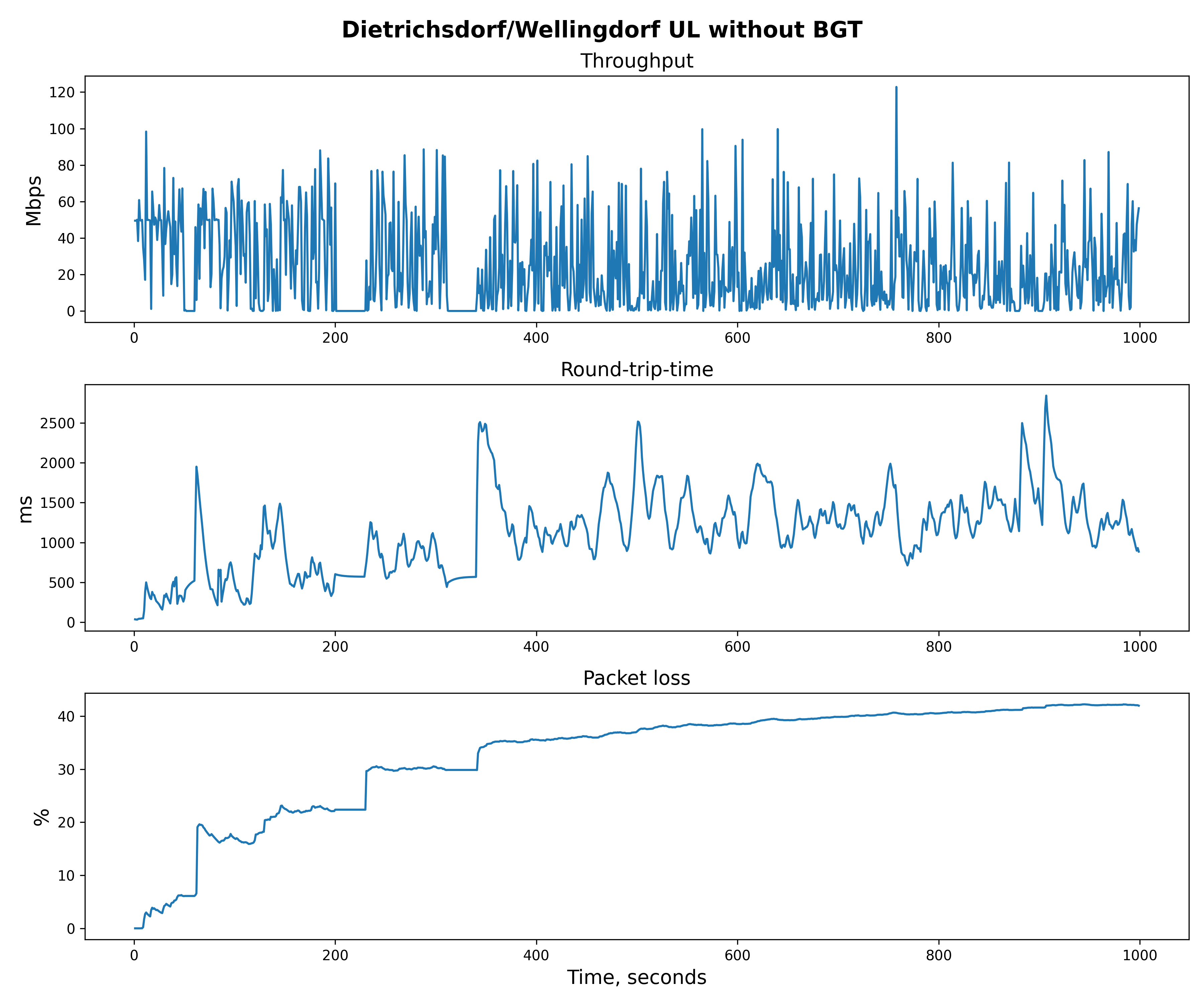}
    \caption{DIT UL run without the background traffic.}
    \label{figure:dit_nt}
\end{figure}
First, the results of running REV and DIT scenarios for the UL direction with and without background traffic are compared. The plots containing the throughput, RTT, and PLR are shown in Fig.~\ref{figure:rev_bg}, Fig.~\ref{figure:dit_bg}, Fig.~\ref{figure:rev_nt}, Fig.~\ref{figure:dit_nt}.

We transmit 50 Mbits of data in each case for 1000 seconds. The background traffic (BGT) is modeled by BackgroundUE modules implemented in the simu5g OMNeT++ library~\cite{simu5g}. The number of UE devices is randomized and depends on the particular run: we populate each carrier with a random number of background units by default ranging from 10 to 20, and they produce both downlink and uplink traffic by default ranging from 300 Kbps to 1.5 Mbps. The amount of downlink and uplink traffic can be set with the special parameter during network initialization.

Looking at the plots for UL scenarios, the following conclusions can be drawn:
\begin{itemize}
    \item[\textcolor{green}{\checkmark}] 5G network can provide 50 Mbps under a light workload, but when there is enough background traffic, then the realistic bandwidth can shrink up to 20-30 Mbps for the REV and 10-20 for the DIT and it jumps from low to high values. This was observed many times during the real runs, causing e.g., a \textbf{stalling effect} for the video streams.
    \item[\textcolor{green}{\checkmark}] With BGT, when the traffic starts to get lost, the RTT stays on the level of a second for the REV and 2 seconds for the DIT. However, the RTT is a round trip, and the one-way latency is usually assumed to be half of the RTT. Note that if the amount of traffic sent passes the available bandwidth, the RTT stays around 20-50 milliseconds (e.g., from 0 to about 200 seconds at the REV plots), which is the typical range for 5G networks.
    \item[\textcolor{green}{\checkmark}] Sailing at high speed from the left coast to the right coast requires adaptive communication because the network condition along the way is very unstable. During the docking, the connection is more or less stable.
\end{itemize}

We also explain in detail for these plots all network events that occurred during the run and how they affect the metrics in general:
\begin{itemize}
    \item[\textcolor{blue}{\textbullet}] The first reason for the "jumpy" throughput behavior is \textbf{long distance from the base station}. Since we had no direct information about the scheduling discipline of Vodafone 5G base stations, we assumed that it is proportional fair (PF) in UL and DL directions~\cite{scheduler}. This means that the scheduler does not additionally down-prioritize connected UEs that are far away, but the signal power becomes exponentially worse anyway. This causes bandwidth degradation in the BGT scenario.
    \item[\textcolor{blue}{\textbullet}] Another reason is the \textbf{handovers}. A handover is a process of changing the base station. We did not have the information about the handover management from Vodafone, so we assumed it was handled as usual by measuring the channel quality indicator (CQI) feedback probes called Channel State Information (CSI) Report~\cite{csi}. A handover causes temporary jumps in RTT and an increase in the global PLR in both plots. This is because during a handover, the connection is lost for some time as two base stations establish the handover. However, the data is stored in 5G buffers, so it will be delivered later. This explains why the throughput sometimes jumps above 50 Mbps after a drop. When the ship is moving at high speed, the handovers could occur very often.
    \item[\textcolor{blue}{\textbullet}] The overall "sinusoidal" behavior of throughput and the "jumpy" behavior of RTT and PLR could be characterized as \textbf{network congestion}. Network congestion occurs when traffic exceeds the capacity of network resources, creating a bottleneck where data packets accumulate faster than they can be processed or forwarded. Buffers within network devices become overloaded, resulting in packet loss as incoming packets are dropped due to a lack of buffer space. The following chapters discuss how to detect and deal with congestion.
\end{itemize}

\subsubsection{Downlink}
\begin{figure}[!t]
    \includegraphics[width=\textwidth, height = 6.8cm]{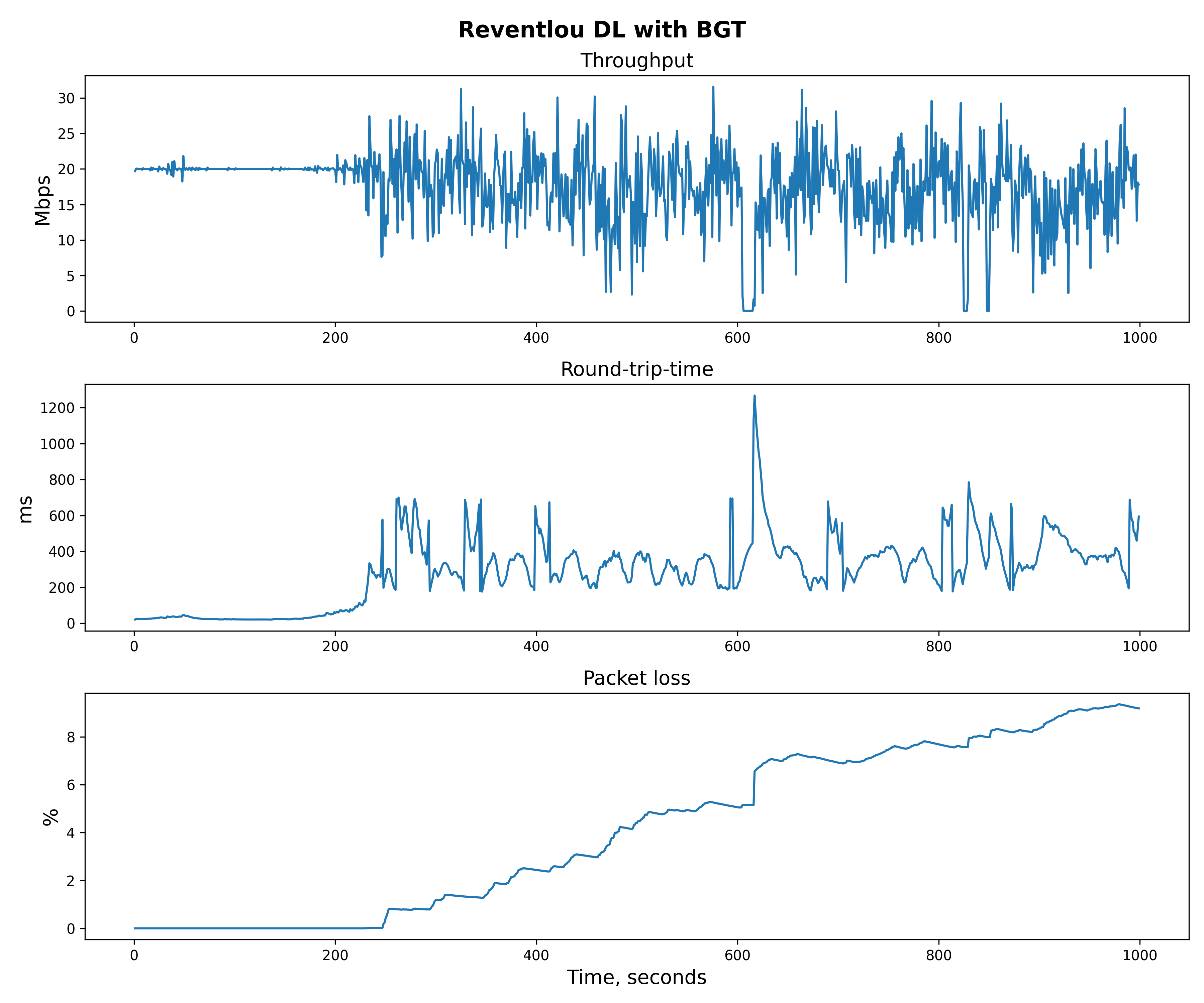}
    \caption{REV DL run with the background traffic. Here, the LiDAR sensor mounted on the Reventlou pier is modeled. It transmits 20 Mbps to the ship, defining a downlink scenario.}
    \label{figure:rev_dl}
\end{figure}

The results for the REV downlink scenario are shown in Fig.~\ref{figure:rev_dl}. Here we can see that the throughput is more or less stable, although it sometimes oscillates around 20 Mbps due to the high-speed handovers. The RTT stays around 200 ms, sometimes jumping to 600 during the handovers. The big peak in RTT and PLR and a corresponding drop in throughput around 650 seconds on a graph corresponds to the edge case already seen with an \textbf{almost out-of-range} effect corresponding to the fact that the ship is in the middle of the Kiel Bay sailing over the Schwentine route. However, the realistic requirements for this scenario should assume that as long as the ship moves away to a certain distance, it does not need to transmit this stream since it leaves the docking area.

Based on the experiments with the DL scenario, the following conclusions can be drawn:
\begin{itemize}
    \item[\textcolor{green}{\checkmark}] DL provides better throughput, RTT, and lower PLR compared to UL. The ratio has been measured to be about 3 to 1 in simulation experiments correlating with the latest state-of-the-art research~\cite{5g_ul_dl}.
    \item[\textcolor{green}{\checkmark}] Intelligent pier enables \textbf{reliable transmission} with the accepted latency of the streams from the pier, thus assisting with docking maneuvers for all passing ships.
\end{itemize}

\subsubsection{Multi-Homing}

The last scenario is related to testing the multi-homing capabilities in REV. This is when we distribute the traffic over more than one communication channel in parallel. For this purpose, we configured the coastline with the real access points made by Addix (see Fig.~\ref{figure:sim_scen}). We transmit 25 Mbits of data over 5 minutes, including the docking at Reventlou and full speed towards the Dietrichsdorf station. The data is split equally between both communication stacks (i.e. 12.5 Mbps each). We also use only valid public Wi-Fi frequencies, namely 2.4 and 5 GHz.

\begin{figure}[!t]
    \includegraphics[width=\textwidth, height = 6.5cm]{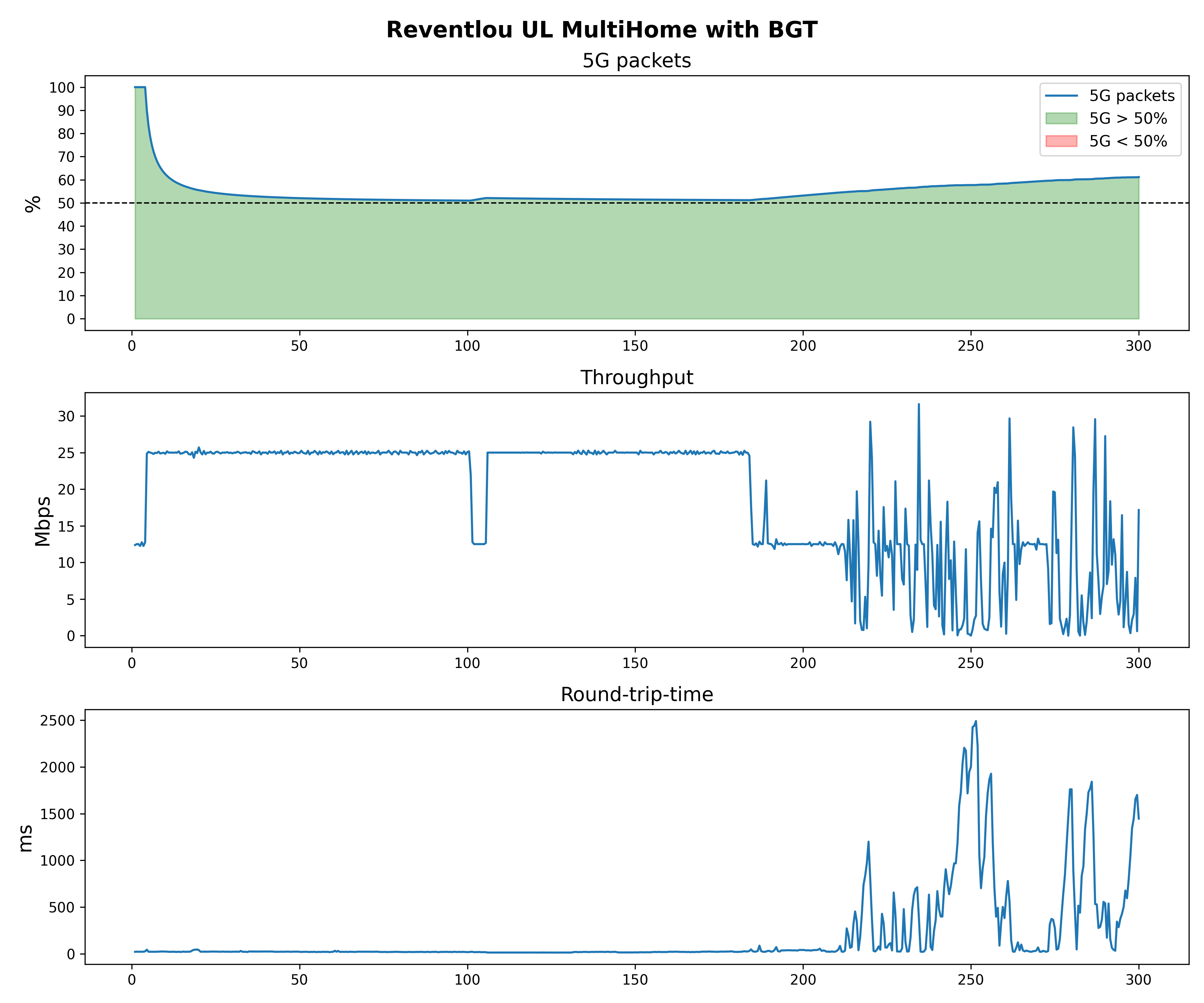}
    \caption{Multi-Homing REV run with the background traffic.}
    \label{figure:rev_multi}
\end{figure}

The results are shown in Fig.~\ref{figure:rev_multi}. It can be observed that the range of the Wi-Fi signal is quite small: after 210 seconds it was lost and half of the traffic was not sent. This time step corresponds to the \textbf{range of Wi-Fi for about 80 meters from the coast}. This means that the Wi-Fi signal could be useful for docking, but it could not be fully reliable even in the range of coverage, because at about 105 seconds there was a handover where the Wi-Fi signal broke and disappeared for 2-3 seconds. After losing the Wi-Fi signal, 5G overtakes the rest of the traffic, but also experiences problems because we modeled 2 times more background traffic than in UL REV/DIT scenarios due to the additional interference. We did not compare bandwidth, signal quality, etc. between wireless and cellular stacks, as it is known from the literature that 5G generally outperforms Wi-Fi version 5 and 6~\cite{5g_wifi}. The main goal was to investigate the range of the Wi-Fi signal along the coast and the overall connection within the increased signal-dense area.

Wi-Fi and LTE/NR coexistence in unlicensed spectrum is a popular topic in research now. Since public Wi-Fi and 4/5G networks operate in strictly defined frequencies, it causes disruptive signal interference and out-of-order packets. That's why there are, e.g., no smartphones that implement this coexistence by splitting traffic between both communication channels. However, in the non-licensed spectrum, the results of this coexistence could be very beneficial~\cite{5g_wifi_co}.

Based on the multi-home scenario experiments, the following conclusions can be drawn:
\begin{itemize}
    \item[\textcolor{green}{\checkmark}] The multi-homing of Wi-Fi and 4G/5G stacks could be beneficial, but \textbf{the most prioritized data should be transmitted over 5G, even if 5G is non-standalone}. The data management system should always take into account the distance from the coast to redirect the traffic over the cellular stack.
    \item[\textcolor{green}{\checkmark}] Multi-homing creates the problem of \textbf{out-of-order packets}. Special jitter buffers and additional checks on timestamps and packet flags must be used to synchronize data collection from both communication channels. 
    \item[\textcolor{green}{\checkmark}] If the stability of the Wi-Fi signal along the coast is at a high level (no long network interruption during handover), then it could help to increase the amount of data that can be transferred while maintaining the same level of reliability. 
\end{itemize}

\subsection{Summary}
Evaluating the quality of data transmission involves analyzing three primary metrics: throughput, round-trip time, and packet loss. Throughput measures the actual useful data received at the application layer, while round trip time refers to the time delay between the sender and receiver for a round trip. Packet loss is the percentage of data packets that do not reach their destination.

Two docking scenarios with docking at the stations Reventlou (REV), Dietrichsdorf and Wellingsdorf (DIT) are simulated in the Gymir5G framework developed by the "Intelligent Systems" working group for the "Schwentine" area of the Kiel Bay. These scenarios model different network conditions and deployment scenarios to evaluate the data transmission performance.

Summarized conclusions:
\begin{itemize}
    \item \textbf{Uplink}. Detailed results are provided for uplink (UL) scenarios in both REV and DIT, taking into account various factors such as background traffic. Throughput, RTT, and PLR behavior are analyzed, highlighting the impact of network congestion, handovers, and signal strength on data transmission performance:
    \begin{itemize}
        \item[\textcolor{green}{\checkmark}] In the REV scenario, the ship shows the ability to more or less stably transmit 50 Mbps of data to shore at high speed under good workload conditions (around 80 Mbps in the most ideal case) but drops to 20-30 Mbps in the middle of Kiel Bay, which can cause real video streams to stall.
        \item[\textcolor{green}{\checkmark}] RTT remained in the 20-50 millisecond range when bandwidth was sufficient, but reached up to 1 second and rarely even more during handovers and out-of-coverage events.
        \item[\textcolor{green}{\checkmark}] Network congestion, long distances from base stations, and frequent handovers were identified as key factors affecting transmission performance.
        \item[\textcolor{green}{\checkmark}] The DIT scenario showed similar trends, but with lower throughput (10-30 Mbps in comparison to 40-50 Mbps) due to non-standalone 5G base stations and higher background traffic from the nearby Kiel University of Applied Sciences.
    \end{itemize}
    \item \textbf{Downlink}. The downlink (DL) scenario for REV is examined, showing stable throughput and the role of an intelligent pier in ensuring reliable data transmission during docking maneuvers:
    \begin{itemize}
        \item[\textcolor{green}{\checkmark}] Throughput remained stable at approximately 20 Mbps (two LiDAR streams), with occasional fluctuations due to high-speed handovers. 
        \item[\textcolor{green}{\checkmark}] RTT stayed around 200 ms, jumping to 600 ms in edge cases where the ship was out of range.
        \item[\textcolor{green}{\checkmark}] Packet loss remains below the acceptable rate of 1-3 \% at normal distance from the sensor.
    \end{itemize}
    \item \textbf{Multi-homing}. The multi-homing scenario in REV is explored, focusing on the simultaneous use of Wi-Fi and 5G communication stacks:
    \begin{itemize}
        \item[\textcolor{green}{\checkmark}] Results show challenges with Wi-Fi signal range limitations and issues related to out-of-order packets during data transmission requiring explicit synchronization.
        \item[\textcolor{green}{\checkmark}] The Wi-Fi signal was lost approximately 80 meters from the shore, affecting the reliability of the connection during docking.
        \item[\textcolor{green}{\checkmark}] While multi-homing offers potential benefits and an increase in bandwidth, prioritizing 5G for critical data transmission is critical to ensure reliable and efficient communications.
    \end{itemize}
\end{itemize}

\clearpage
\section{Real-time media streaming}
\subsection{Types of data and their transmission}
In terms of transmission, there are two types of data: "live" and "non-live."\textbf{Live or real-time data} transmission differs from non-live in several ways. Live data transfer involves the continuous and immediate delivery of data from source to destination in real time, without significant delays or the possibility of buffering. This is critical for applications where timely and synchronized delivery is essential, such as live video streaming. In contrast, \textbf{non-live data} transmission refers to the asynchronous delivery of data where the data is sent and received independently of real-time events or interactions, such as serialized file (media) consumption. While non-live data does not pose significant problems due to the lack of latency requirements, live data delivery needs to be optimized and carefully controlled. The main example of live data for the "Wavelab" ferry is the video streams that transmit video from the cameras installed throughout the unit to the shore control center.

In video transmission, the choice between UDP (User Datagram Protocol) and TCP (Transmission Control Protocol), the two basic transport layer protocols, has long been a trade-off between speed and reliability. UDP prioritizes speed by transmitting data packets without establishing a connection or verifying delivery, making it well-suited for applications where real-time delivery is critical. Unlike TCP, which ensures reliable delivery through mechanisms such as acknowledgement and retransmission, UDP sacrifices reliability for lower latency and faster transmission, making it the preferred choice for video delivery in scenarios where occasional packet loss is acceptable.

However, modern streaming protocols have emerged to address the shortcomings of UDP and TCP and provide a balance between latency, self-regulation, and reliability. These protocols focus on optimizing streaming performance while considering factors such as network conditions, device compatibility, and quality of user experience. For the CAPTN Fjord5G project, we emphasize three key aspects:
\begin{enumerate}
    \item \textbf{Latency requirements}, aiming to minimize the delay between video capture and playback to improve real-time interaction and control engagement.
    \item \textbf{Deployment and management complexity}, looking for solutions that are easy to deploy and manage under varying network conditions.
    \item \textbf{Popularity}, considering adoption and support of the protocol by industry players, developers, and end users to ensure widespread acceptance and interoperability.
\end{enumerate}

\begin{figure}[!t]
\centering
    \includegraphics[width=\textwidth]{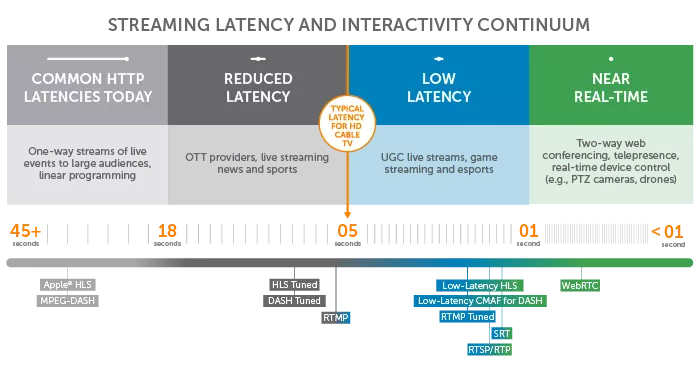}
    \caption{The latency comparison of the modern streaming protocols. Taken from~\cite{wowza}.}
    \label{figure:wowza}
\end{figure}

Among the available streaming protocols, Real-Time Communication for the Web (WebRTC) stands out as the only one that offers sub-second latency, making it a preferred choice for applications that require real-time communication and interaction~\cite{webrtc_def}. The research conducted by the Wowza portal~\cite{wowza} is shown in Fig.~\ref{figure:wowza}. WebRTC accomplishes the sub-second latency by leveraging peer-to-peer (p2p) connection and optimized delivery mechanisms. In addition, its broad support across browsers, platforms, and devices makes it highly accessible and versatile, addressing a wide range of use cases. WebRTC has become the go-to solution for various real-time communication applications, including video conferencing, live streaming, and interactive gaming. In agreement with the industry partners, WebRTC was selected as the main streaming protocol for the CAPTN Fjord5G research project.

\subsection{WebRTC protocol}\label{section:3.2_webrtc_protocol}
WebRTC is a versatile and widely used peer-to-peer protocol for real-time communication over the Internet, enabling applications such as video conferencing, surveillance, live streaming, etc. At its core, WebRTC relies on two fundamental media transport components: the Real-time Transport Protocol (RTP) and the Real-time Transport Control Protocol (RTCP). RTP handles the actual transmission of data, including audio and video streams. It segments media into packets and adds timestamp and sequence number information. At the same time, RTCP serves as a control protocol for monitoring network conditions, providing feedback on packet loss, inter-arrival deltas, jitter, and round-trip time, as well as synchronizing different timestamps within an established connection. 

When dealing with video streams in WebRTC, the sender processes the video data before transmitting it. This processing involves video encoding techniques to transform raw video data into segmented frames of reduced size, followed by payloding these frames into RTP packets and then sending them via pacer to prevent the creation of large bursts. This topic will be discussed in detail in the next chapter. At the receiving end, the incoming RTP packets are processed to reconstruct the frames. This involves reordering the packets based on sequence numbers, which are held in a jitter buffer. The reconstructed video is then rendered to the user.

In addition to its media delivery capabilities, the WebRTC protocol also provides the ability to stream non-live data over \textbf{data channels}~\cite{webrtc_dc}. These data channels are built on top of the Stream Control Transmission Protocol (SCTP), which provides features such as message-oriented communication and multihoming support. It allows applications to exchange files, share documents, and even synchronize application states. Unlike traditional HTTP-based protocols optimized for bulk data transfer, WebRTC data channels are also designed for low-latency and bi-directional communication. Leveraging the same underlying infrastructure as WebRTC's media delivery, data channels inherit the protocol's sub-second latency and peer-to-peer connectivity while enabling reliable non-live data transfer. 

\begin{figure}[!b]
\centering
    \includegraphics[width=\textwidth, height=5.8cm]{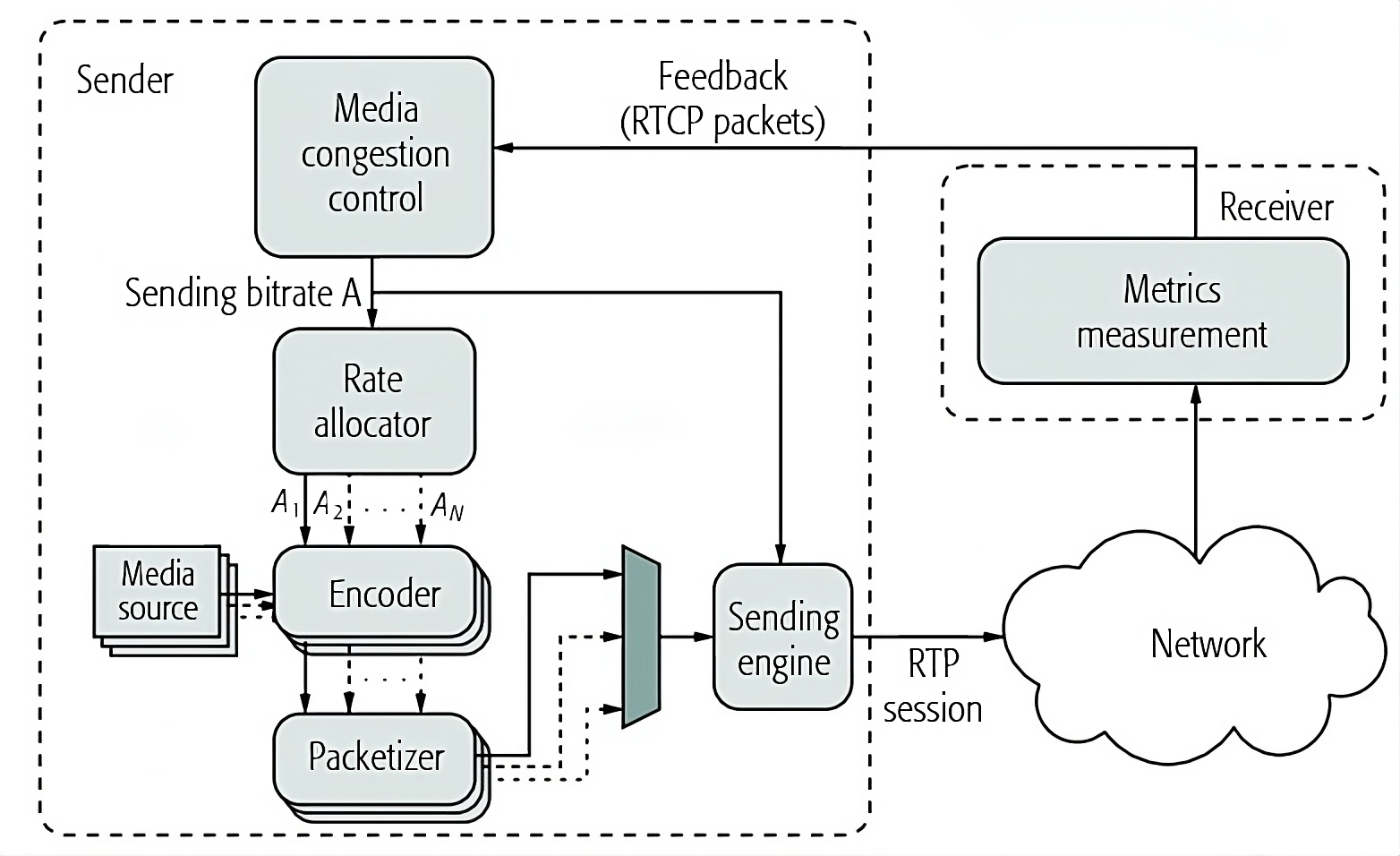}
    \caption{WebRTC media processing scheme. Taken from~\cite{webrtc_congestion_control}.}
    \label{figure:webrtc_congestion_control}
\end{figure}

WebRTC also includes congestion control mechanisms, the scheme of which is shown in Fig.~\ref{figure:webrtc_congestion_control}. This involves adjusting the video bitrate in response to estimated network bandwidth and can be done either on both the sender and receiver sides or only on the sender side. The first approach, which relies on the receiver periodically sending bandwidth estimates, is now considered deprecated to avoid overloading the client, which is usually a web browser. The second approach, known as Transport-Wide Congestion Control (TWCC), uses periodic feedback packets containing the inter-arrival delay of each received packet~\cite{twcc}. This gives the sender information about packet loss and arrival deltas and allows them to estimate bandwidth using congestion control algorithms.

\subsection{WebRTC simulation experiments}
\subsubsection{WebRTC features in a simulation}
Since WebRTC was selected as the primary streaming protocol and there is currently no implementation of WebRTC in OMNeT++, several key features have been implemented, each contributing to a more comprehensive understanding of real-time communication scenarios. Gymir5G encompasses the following mechanisms:
\begin{enumerate}
    \item \textbf{Negative Acknowledgment} (NACK) and packet retransmission. It allows the receiver to notify the sender of missing packets so that the sender can retransmit them. A user can configure the maximum age of the retransmission buffer at the sender, the maximum number of NACKs for a single packet, the maximum wait time for retransmission, and to resume playback at the receiver application. 
    \item \textbf{Forward Error Correction} (FEC). FEC adds redundancy, allowing missing packets to be reconstructed without requesting retransmission. Gymir5G simulates the very basic version of Uneven Level Protection (ULP)~\cite{ulpfec}, i.e., packets with a keyframe are better protected. A user can manage the frequency of FEC injection at the sender and the frequency of repair attempts at the receiver application.
    \item \textbf{Playback/jitter buffer}. This is a simple buffer on the receiver side that collects all incoming packets and waits with some maximum delay for lost packets to continue playback. It does not perform any decoding mechanisms or apply any jitter reduction techniques.
    \item \textbf{Sender and Receiver Reports} (SR, RR) and \textbf{Transport-Wide Congestion Control} (TWCC) packets. SR and RR packets are sent by each side in WebRTC to provide feedback on network performance. They are an integral part of the Real-Time Transport Control Protocol (RTCP), which works alongside RTP within WebRTC~\cite{rtcp}. TWCC feedback plays a critical role in congestion control by providing reports on the inter-arrival delays of all received packets~\cite{twcc}. A user can set the periods for all of the above feedback and enable "no-cost" delivery: it guarantees that the feedback won't be lost and will be delivered with zero latency.
\end{enumerate}

\subsubsection{Sandbox scenario}\label{section:3.2.2_sandbox}
\begin{figure}[!t]
\centering
\includegraphics[width=0.85\columnwidth, height=5.8 cm]{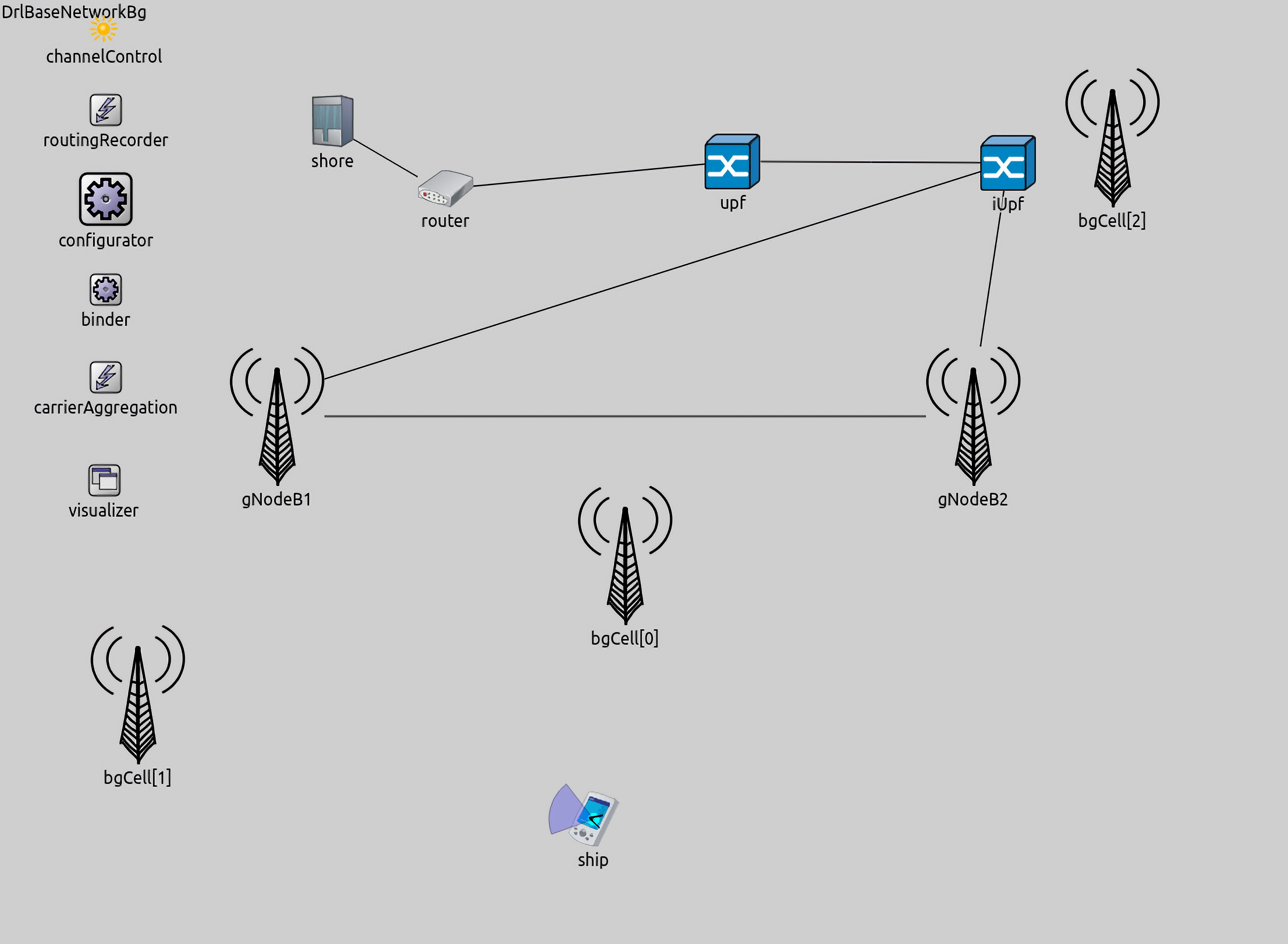}
\caption{Gymir5G network for the "sandbox" scenario.} 
\label{figure:simu5g_sandbox}
\end{figure}

For the experiments with the WebRTC protocol, we have developed a special "sandbox"-like scenario where a device transmits the data to a server within the 5G network~\cite{me_arcs}. The setup is adapted to the maritime domain of the CAPTN Fjord5G project: a device is a ferry sending the video data, and the receiver is a land-based station. The other 5G devices are distributed over the map. There are also background cells that mimic other vendors' 5G base stations in the area, causing signal interference and noise. A schematic representation of a sandbox scenario in Gymir5G is shown in Fig.~\ref{figure:simu5g_sandbox}.

The idea of the presented sandbox scenario is to provide different network experiences, from almost perfect to loss rates of over 70\%, by increasing the number of devices and their data rates, while limiting the areas where the agent moves during the run. When it moves to the lower and right lower areas, it suffers from severe congestion. On the contrary, it enjoys perfect 5G coverage, minimal delays, and scheduling priority in the top and top-left areas. The repeated OMNeT++ seeds are used to ensure the reproducibility of a simulation process. For the same reason, there are no random elements in the ferry's paths and no bitrate oscillations in the streams.

As in the previous section, we developed two main configurations for the above scenario: "easy" and "congested". "Easy" represents the unproblematic network with good coverage, where there could be little loss and almost no congestion. "Congested", on the other hand, represents the network with some periodic congestion episodes that can last from 1-2 to 10-20 seconds, resulting in higher latency, packet loss, and bandwidth saturation. The ferry transmits a 4K simulated video stream in the UL direction with a 10 Mbps bitrate.

The following use cases are examined and evaluated for both scenarios in the following subsection:
\begin{itemize}
    \item Simple streams over UDP and TCP transport protocols.
    \item WebRTC streams with realistic and stronger NACK settings.
    \item WebRTC streams with and without FEC support.
\end{itemize}

\subsubsection{Results}
\paragraph{UDP vs. TCP}
WebRTC can be implemented over both UDP or TCP transport protocols. When running over UDP, WebRTC benefits from \textbf{lower latency and reduced overhead} due to UDP's connectionless nature. However, UDP does not guarantee the delivery or order of packets, which can lead to packet loss, requiring additional mechanisms such as retransmission and congestion control to ensure reliability. On the other hand, running WebRTC over TCP provides \textbf{reliability and orderly delivery of packets}. However, the inherently connection-oriented nature of TCP can introduce higher latency and increased overhead.

\begin{figure}[!b]
\centering
    \includegraphics[width=\textwidth, height=6.8cm]{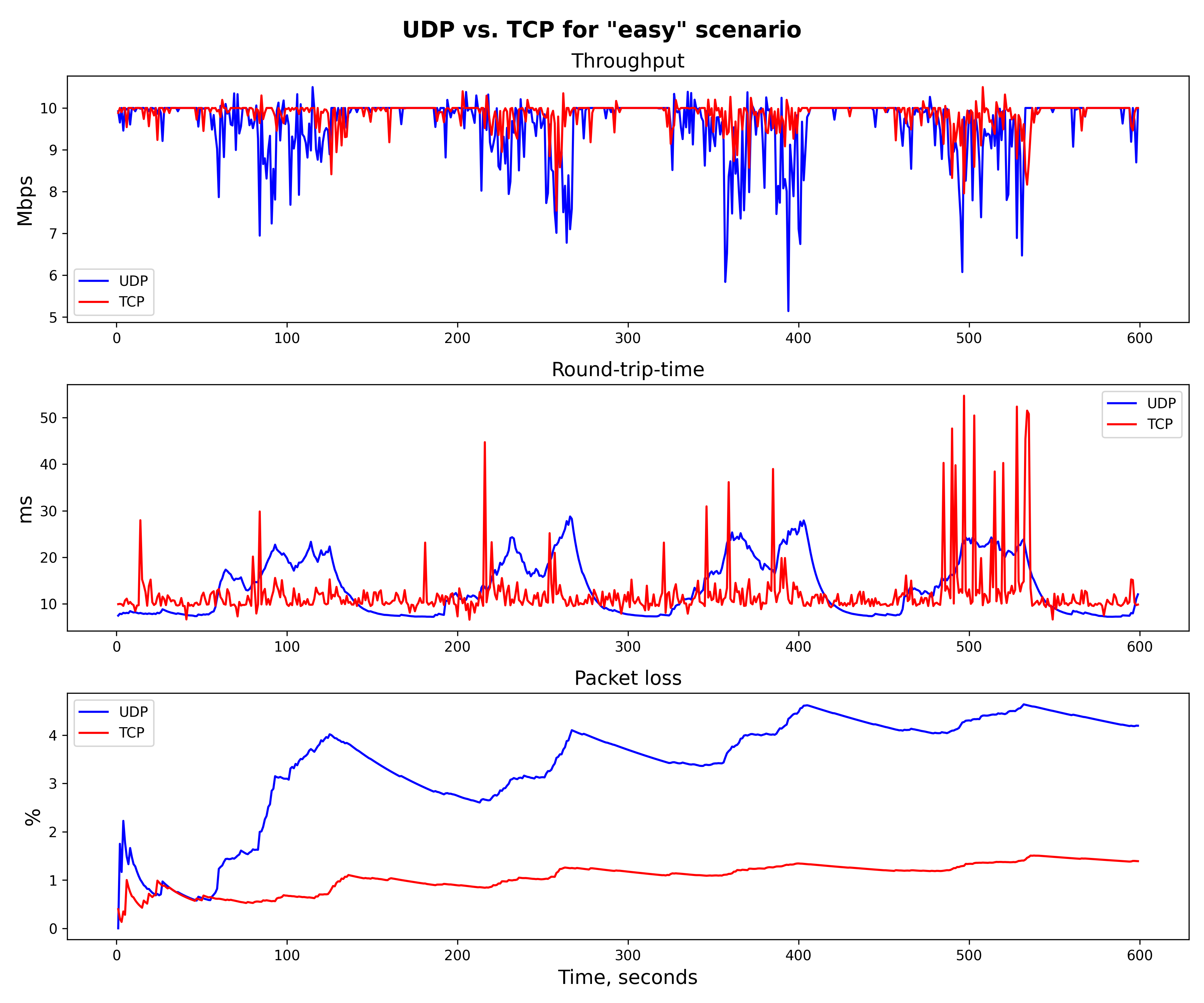}
    \caption{The comparison between UDP and TCP transport protocols for the "easy" scenario.}
    \label{figure:w_udptcp_easy}
\end{figure}
\begin{figure}[!b]
\centering
    \includegraphics[width=\textwidth, height=6.8cm]{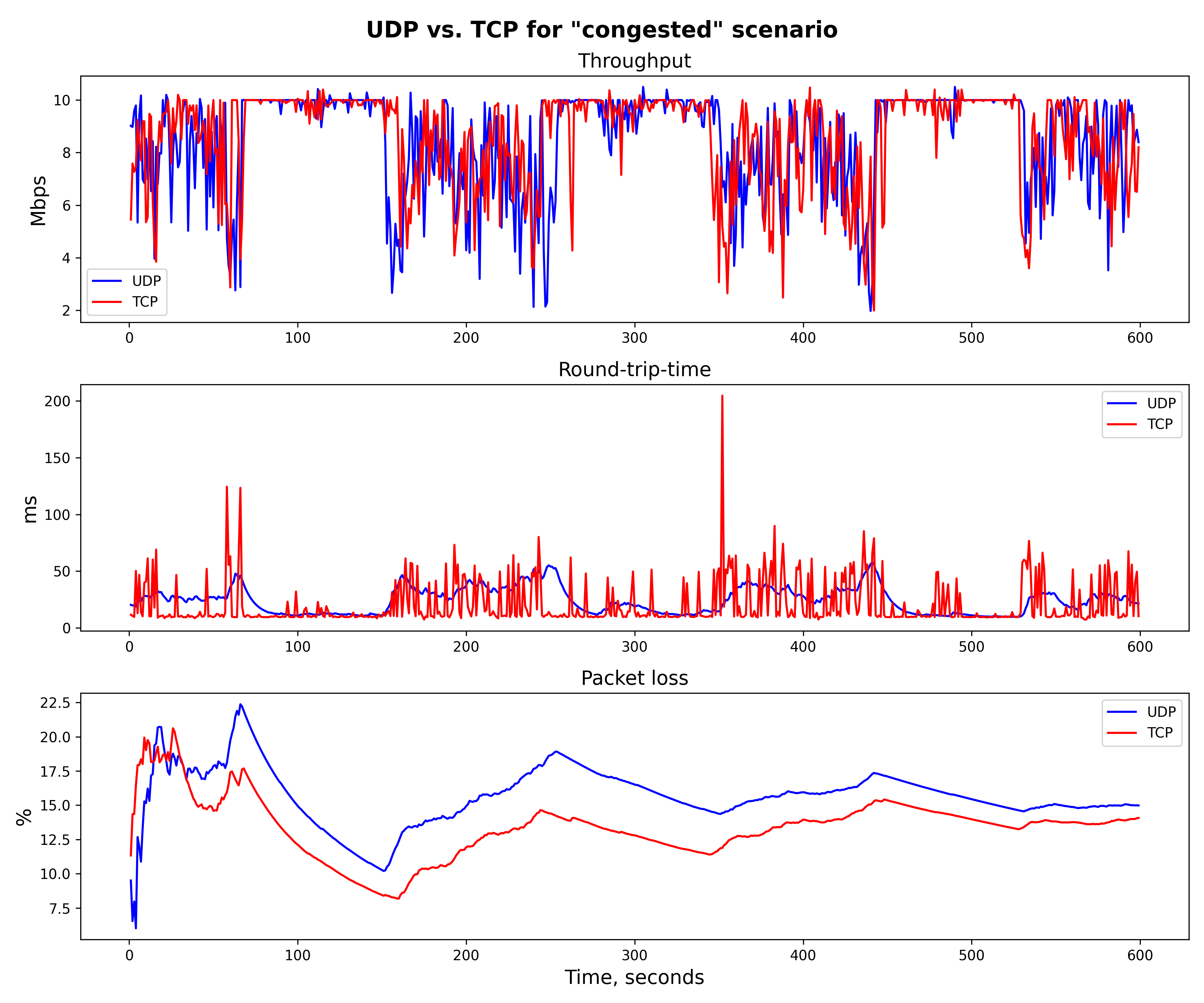}
    \caption{The comparison between UDP and TCP transport protocols for the "congested" scenario.}
    \label{figure:w_udptcp_hard}
\end{figure}

We ran two experiments for "easy" and "congested" scenarios comparing the behavior of UDP and TCP streams, see Fig.~\ref{figure:w_udptcp_easy} and Fig.~\ref{figure:w_udptcp_hard}. They demonstrate the above: when there are only small congestions, TCP is more effective in coping with the PLR and has a more stable rate, but even in such conditions it produces a higher RTT than UDP. On the other hand, in the "congested" scenario, the difference in PLR and throughput is not as large. However, the RTT for TCP is significantly higher. This RTT overhead is much worse in the case of multiple streams and is known as \textbf{Head-of-Line (HOL) blocking}: until the lost packet for a particular stream is not acknowledged or considered finally lost, all packets in a queue for that and other streams remain unprocessed~\cite{hol}. This also explains why the PLR for TCP is so large: the jitter buffer at the receiver side has a certain playout delay, and when it is exceeded, it marks the packet as "lost". When the HOL is further unblocked, the incoming packet burst is not fully decoded: only the most recent ones are decoded, while older ones are not, because the whole burst takes too much time for sequential decoding, and it is crucial to start showing something after the long video stall caused by the HOL blocking. The "congested" scenario results fully reflect why WebRTC is often run over the UDP transport protocol rather than TCP in real-world applications.

\paragraph{NACK}
\begin{figure}[!b]
\centering
    \includegraphics[width=\textwidth, height=6.8cm]{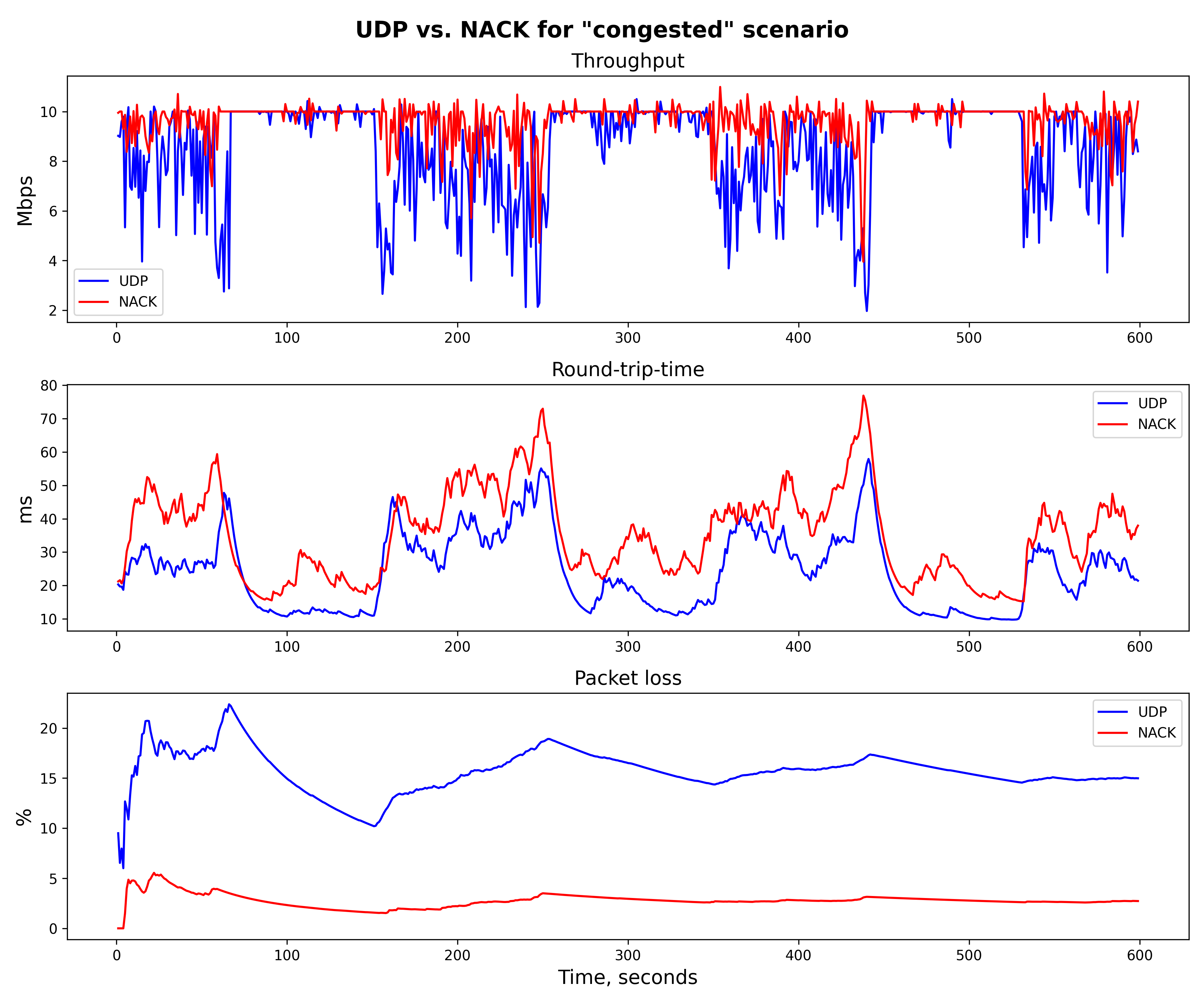}
    \caption{The comparison between UDP and WebRTC+NACK protocols for the "congested" scenario.}
    \label{figure:w_udpnack_hard}
\end{figure}

From the previous paragraph, it is clear why UDP is generally considered more suitable for RTC than TCP. However, its unreliability is a serious issue that is addressed in WebRTC by introducing the retransmission mechanism by sending Negative Acknowledgement (NACK) requests. It differs from TCP's ACKs in the word "negative": only missed packets are requested to be retransmitted, rather than blocking transmission until a positive acknowledgement of a successfully delivered packet arrives. Fig.~\ref{figure:w_udpnack_hard} illustrates the difference between pure UDP and basic WebRTC with NACK in the "congested" scenario. The NACK mechanism also introduces an overhead to RTT, but it is much smaller than TCP, and more importantly, it is much \textbf{more effective at dealing with lost packets than both UDP and TCP}. NACK packets, and the retransmission mechanism in general, are enabled by default for almost all real-world WebRTC producers and clients.

We used the default parameters for the retransmission mechanism taking the Google Chrome configuration as an example~\cite{nack_chrome}:
\begin{itemize}
    \item[\textcolor{blue}{\textbullet}] Maximum number of NACK requests for a single packet: 10.
    \item[\textcolor{blue}{\textbullet}] Time between consecutive NACK requests: 20 milliseconds.
    \item[\textcolor{blue}{\textbullet}] Maximum packet "age", i.e. how long it can exist in a missing list: a) it is not "older" than 10000 sequence numbers from the last delivered, b) the size of the missing list is not greater than 1000 packets, c) the maximum number of retransmissions for this packet does not exceed 10, d) there is no next full decodable keyframe available.
    \item[\textcolor{blue}{\textbullet}] Sender restrictions: a) avoid retransmissions within the last RTT period, b) limit retransmissions based on channel bandwidth estimation, c) prioritize packets when pacing is enabled.
\end{itemize}

\begin{figure}[!b]
\centering
    \includegraphics[width=\textwidth, height=6.8cm]{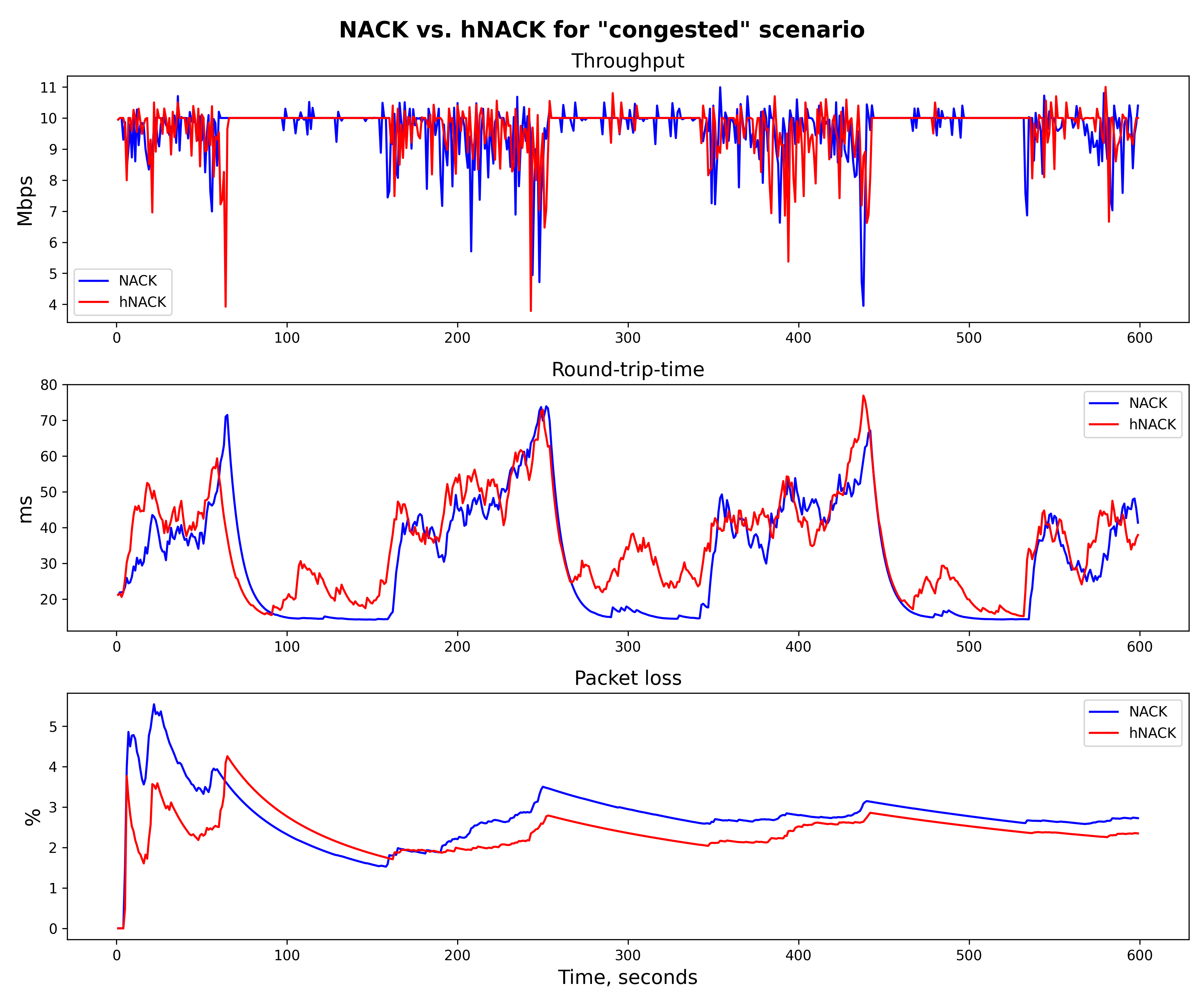}
    \caption{The comparison between NACK and "hard"-NACK WebRTC protocols for the "congested" scenario.}
    \label{figure:w_nackhnack_hard}
\end{figure}

We have simulated these parameters and also two times increased default parameters for WebRTC streams. They are compared on the "congested" scenario and marked in Fig.~\ref{figure:w_nackhnack_hard} as NACK and hard NACK (hNACK) configurations. It can be seen that deviating from the default parameters does not bring significant benefits: it logically makes the RTT a bit higher and brings a small decrease in the PLR. This also explains why most WebRTC peers often do not allow easy configuration of retransmission parameters.

\paragraph{FEC}
Another WebRTC mechanism for mitigating packet loss is Forward Error Correction (FEC). WebRTC uses it as a proactive mechanism by adding redundant information to transmitted data packets. This "redundant information" consists of additional data bits calculated from the original packet payload using error-correcting codes~\cite{ulpfec}. FEC allows receivers to reconstruct lost or corrupted packets without the need for retransmission. This preemptive approach reduces packet loss and ensures smoother playback for RTC. It is particularly beneficial in scenarios where packet loss is common or expected, providing an additional layer of resilience to network restrictions.

\begin{figure}[!b]
\centering
    \includegraphics[width=\textwidth, height=6.8cm]{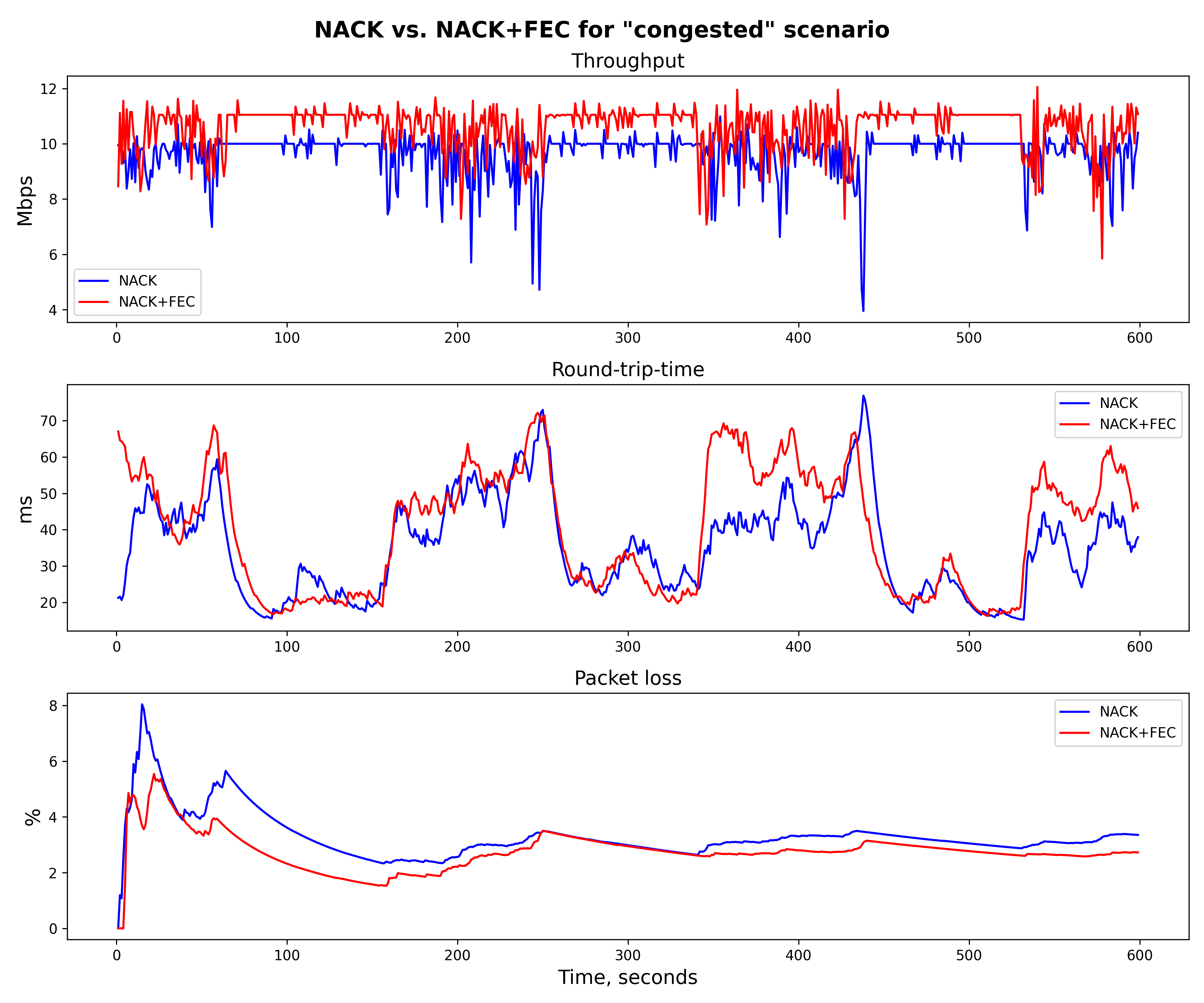}
    \caption{The comparison between the standard WebRTC protocol with NACK and the one with NACK + FEC for the "congested" scenario.}
    \label{figure:w_fec_hard}
\end{figure}

Fig.~\ref{figure:w_fec_hard} compares the WebRTC stream with and without FEC, assuming it uses UDP and NACK is enabled. It shows that while it adds overhead to throughput and RTT, it reduces the PLR. This could be particularly beneficial in the case of uneven protection: stronger protection of keyframes reduces the chance of losing them, which greatly improves the smoothness of playback. As a result, FEC is often used in real-world WebRTC systems to improve overall transmission reliability, especially in applications where low latency and seamless communication are critical.

\subsection{Streaming with WebRTC on the real ferry}
\begin{figure}[!t]
\centering
    \includegraphics[width=0.75\textwidth, height=4.5cm]{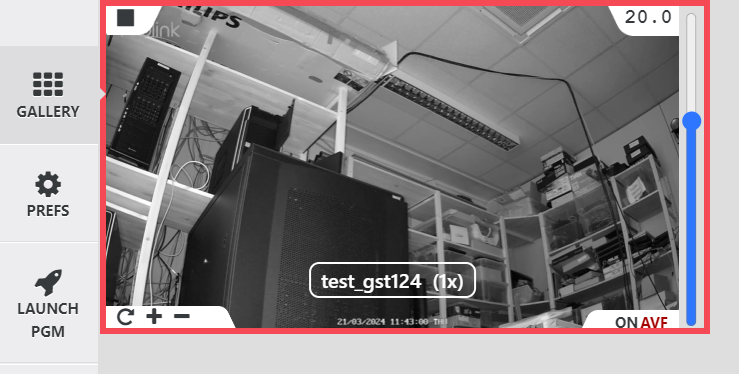}
    \caption{AhoyRTC Director platform with one WebRTC stream.}
    \label{figure:ahoyrtc}
\end{figure}

The AhoyRTC Director platform~\cite{ahoy} is used in the production environment to view the WebRTC streams from the ship by multiple viewers. It is a user interface opened in a browser where multiple WebRTC streams can be turned on and off and also viewed. Fig.~\ref{figure:ahoyrtc} shows a screenshot of the platform. As an example, we streamed a camera in our server room through our app using GStreamer 1.24 as the backend at 4 Mbps at the beginning and 10 Mbps at the end using a hardware-accelerated H264 encoder, in full HD resolution, at 20 fps (shown in the upper right corner). Unfortunately, we cannot show the actual streams from the ship as it is docked in a restricted area most of the time.

\begin{figure}[!b]
\centering
    \includegraphics[width=\textwidth, height=7.5cm]{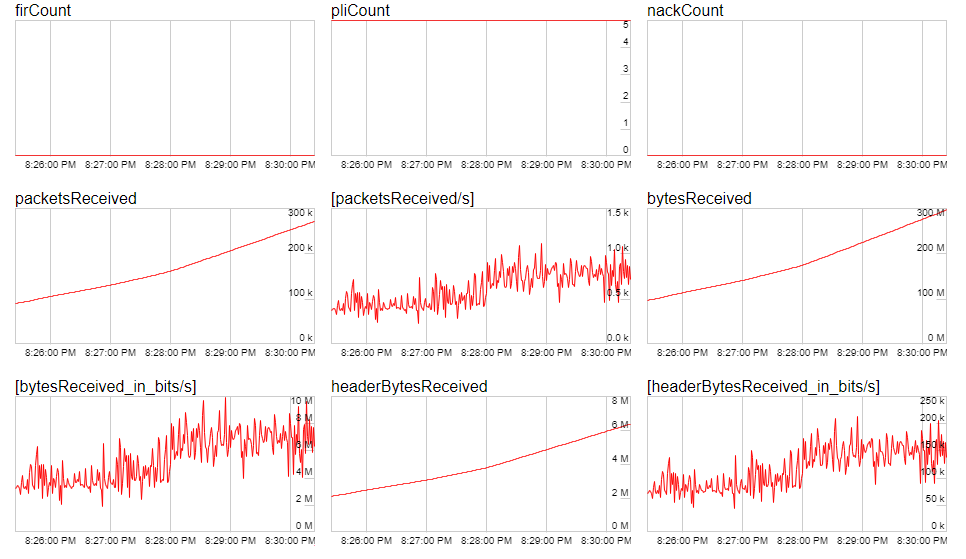}
    \caption{Selected Google Chrome internal WebRTC statistics.}
    \label{figure:chrome_webrtc}
\end{figure}

Using our GStreamer-based "GstWebRTCApp" (see Sec.~\ref{section:1.2_real_testbed}), we receive WebRTC transmission statistics via GStreamer internal callbacks and can publish them in the processed form or use them as feedback for the AI models. However, most of these statistics can be viewed in the Google Chrome browser by calling \textbf{chrome://webrtc-internals}. Fig.~\ref{figure:chrome_webrtc} shows some of these internal statistics that the viewer can follow. The most important is \textit{bytesReceived\_in\_bit/s}, which is simply the received bitrate of the video. Other interesting statistics are \textit{nackCount} and \textit{pliCount}. These are decoding characteristics, among other things: they show the number of NACK and PLI (Picture Loss Indication) requests sent by the viewer's browser. The latter means that the decoder cannot continue decoding until a complete keyframe is delivered. PLI requests often occur during network congestion and are a direct result of packet loss.

\subsection{Summary}
This section discusses various aspects of real-time media streaming, focusing on the transmission of live and non-live data, the choice between UDP and TCP protocols, the emergence of modern streaming protocols such as WebRTC, and experiments conducted to evaluate the performance of WebRTC in simulated scenarios. We also touch on the deployment of WebRTC on the "Wavelab" ferry exploring the AhoyRTC Director platform.

In terms of data transfer, we distinguish between live and non-live data. Live data transfer involves immediate delivery without buffering, which is critical for applications such as live video streaming, while non-live data transfer is asynchronous. The choice of protocol aims to combine the benefits of fast and reliable delivery to stay below sub-second latency. WebRTC was finally chosen as the main streaming protocol for\textbf{ both live and non-live data}: the media data is transmitted directly over the RTP protocol, while the non-live data is transmitted over the data channels using the SCTP protocol. The core components of WebRTC are studied in detail in the text.

Experiments conducted on the Gymir5G simulation platform demonstrate the effectiveness of WebRTC compared to UDP and TCP, particularly in \textbf{mitigating packet loss without notable increase in latency}. Different configurations, including variations in \textit{retransmission} settings and the inclusion of \textit{redundancy} packets, are evaluated and compared to understand their impact on performance. 

Summarized conclusions:
\begin{itemize}
\item [\textcolor{green}{\checkmark}] WebRTC is a modern and versatile protocol suitable for transmitting both real-time live media data and asynchronous non-live data. \textbf{It has been chosen as the main "how" for ship-to-shore sensor data transmission within the CAPTN Fjord5G project}. 
\item [\textcolor{green}{\checkmark}] WebRTC should generally be run over UDP rather than TCP transport protocol to provide the lowest latency for real-time data transfer. However, under "good" network conditions, TCP can offer the advantage of guaranteed delivery without blocking the other streams.
\item [\textcolor{green}{\checkmark}] The WebRTC implementation \textbf{must} support a retransmission mechanism (NACK requests) because it helps mitigate packet loss, and the burst of NACK requests serves as an indicator of network congestion. The default settings are sufficient to provide full functionality, so no additional fine-tuning is required.
\item [\textcolor{green}{\checkmark}] The WebRTC stream \textbf{should} be transmitted with the additional redundant data (FEC packets) in scenarios with a high risk of packet loss. It helps to reconstruct the missing or repair the corrupted packets, although it adds additional redundant throughput. In real life, it could be beneficial in areas without 5G SA coverage, such as the right coast waterways along the Bay of Kiel.
\item [\textcolor{green}{\checkmark}] WebRTC is a widely adopted and easy-to-use protocol that is seamlessly integrated into all major web browsers. It facilitates smooth real-time streaming and allows viewers to monitor stream performance in a simple Google Chrome tab.
\end{itemize}

\clearpage
\section{Data preprocessing: acquisition, encoding, payloading}
To transmit media data over WebRTC, several basic steps must be taken. This process includes data acquisition, compression, encoding, and payloading. For the nautical or weather sensors, this is more straightforward because the data is small and is transmitted over messaging protocols using messaging queue services such as MQ Telemetry Transport (MQTT)~\cite{mqtt}. In this case, all of the above steps are performed by the service and there is no need for intervention or fine-tuning. For media data, this process requires some configuration.

First, data acquisition involves capturing raw information from various sources, such as cameras or LiDARs. After acquisition, the data can be compressed, if necessary, to remove redundant or unneeded information and reduce its size. Encoding techniques are then applied to convert the compressed data into a format suitable for transmission and, more importantly, to control the quality of the data as it is transmitted. Often the compression and encoding stages are combined, especially for video streaming. Finally, the payloading phase involves encapsulating the encoded data into RTP packets (live data) or SCTP packets (non-live data) for transmission over the network. Exploring these steps helps to understand how WebRTC works to make communication smooth and reliable, and how to control it. We take the current information for the CAPTN Fjord5G project at the time of writing (March-April 2024).
\subsection{Acquisition}
\subsubsection{Video acquisition}\label{section:4.1.1_video_acq}
Data acquisition from a camera is a multi-step process in which raw visual information is captured by the camera's sensors, converted into digital data, and made available for further processing. First, light photons are detected by the camera's sensor array, which typically consists of millions of individual pixels. These sensors convert the light intensity and color information into analog electrical signals. Analog-to-digital converters (ADCs) then digitize these signals and convert them into digital data representing the intensity and color of each pixel. The resulting digital data is processed within the camera to make various adjustments, such as white balance, exposure, and noise reduction. Finally, the data is made available through network interfaces so that it can be accessed by any streaming software, such as GStreamer~\cite{gstreamer}.

In the CAPTN Fjord5G project, we use AXIS cameras: P-Series IP cameras and Q-Series thermal cameras\footnote{Datasheets: \textbf{AXIS P1455-LE} \url{https://www.axis.com/dam/public/d8/b3/fd/datasheet-axis-p1455-le-network-camera-en-US-352221.pdf}, \textbf{AXIS Q8642-E PT} \url{https://www.axis.com/dam/public/46/b1/34/datasheet-axis-q8642\%E2\%80\%93e-pt-thermal-network-camera-en-US-359697.pdf}, last accessed: \today}. They are installed on different sides of the ship to provide a complete picture for situational awareness.

There are two ways to acquire video from AXIS cameras:
\begin{itemize}
    \item [\textcolor{blue}{\textbullet}] \textit{Accessing raw data from the camera itself}. By accessing the camera's sensor data directly, one can bypass the compression and processing stages and obtain raw video data in Motion JPEG (MJPEG) or Moving Picture Experts Group-4 (MPEG-4) formats. The raw data is typically 2-3 times the size of the encoded data, and one needs direct access to the device to get the MJPEG/MPEG-4 stream from it, which can be a bottleneck if too many consumers want to capture it simultaneously. But as an advantage, it gives access to the raw data and does not introduce any latency.
    \item [\textcolor{blue}{\textbullet}] \textit{Transmitting the data to some relay}. Alternatively, the camera's API can stream the video data in a raw or an encoded format (e.g., H.264 or H.265, see Sec.~\ref{section:4.2.2_video_codec} for more details) to a relay server or cloud-based service via various streaming protocols that control data delivery, the most popular is Real-Time Streaming Protocol (RTSP)~\cite{rtsp}. This method allows for efficient transmission of video data over a network, enabling remote access to the video stream. However, it introduces latency and may require additional bandwidth, depending on the encoding settings and network conditions.
\end{itemize}

Since AhoyRTC Director assumes a large number of potential viewers, the second option is used primarily. The relay is the RTSPtoWebRTC intermediate service implemented in Go~\cite{rtsptoweb}. It takes the incoming RTSP streams and outputs the same stream without compression or additional encoding into various streaming protocols, including raw RTSP and WebRTC. The latter is used to stream the video directly from the "Wavelab" to the AhoyRTC platform. GstWebRTCApp consumes the relayed RTSP stream and \textbf{provides real-time control over video quality}.

\subsubsection{Point clouds acquisition}
\begin{figure}[!t]
\centering
    \includegraphics[width=\textwidth, height=6cm]{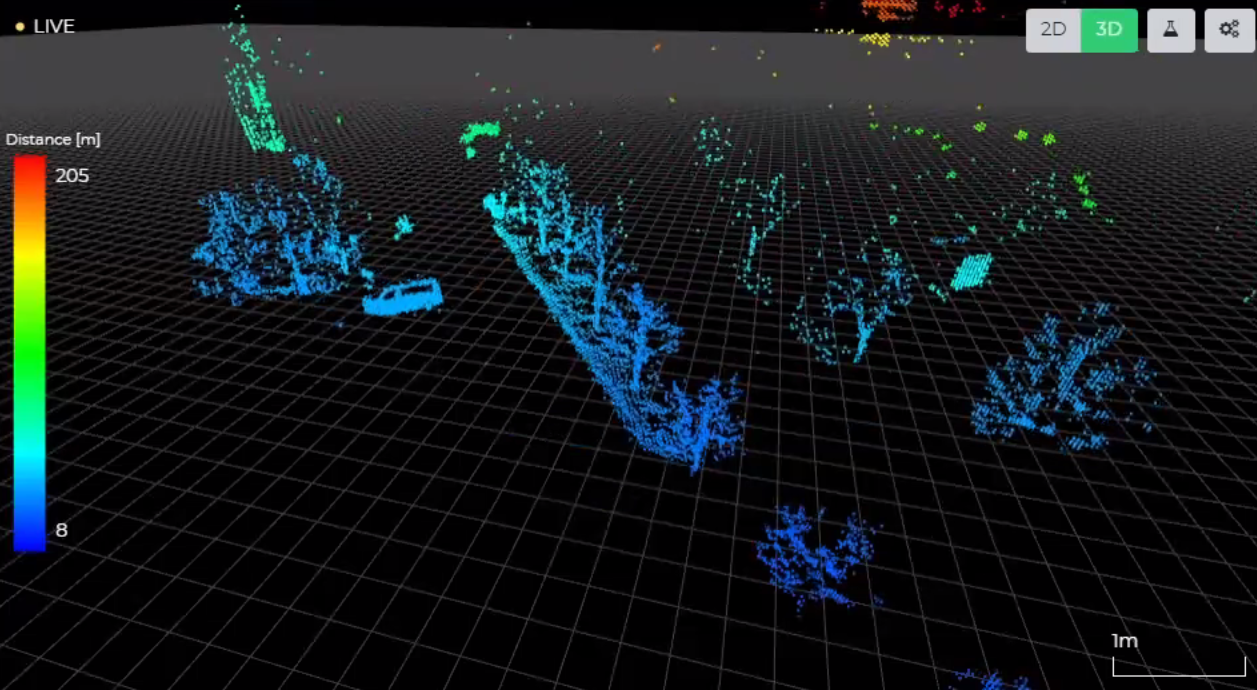}
    \caption{An example of a Velodyne LiDAR snapshot.}
    \label{figure:lidar}
\end{figure}

\label{{section:4.1.2_data_acq_point_clouds}}
Acquiring raw data from a LiDAR sensor involves several steps. First, the LiDAR device emits laser pulses in multiple directions, covering a 360-degree field of view. These pulses bounce off objects in the environment and are reflected to the LiDAR sensor. The sensor then measures the time it takes for each pulse to return, along with the angle at which it is emitted. This information, combined with the known position and orientation of the LiDAR device, allows precise 3D coordinates to be calculated for each point in the scene. The raw sensor output therefore consists of a large number of individual point clouds, with each point representing a specific location in the environment along with additional attributes such as intensity or reflectivity. These point clouds provide a detailed and accurate representation of the environment, capturing both the geometry and characteristics of objects within the LiDAR's range. Fig.~\ref{figure:lidar} shows an example of point clouds.

We analyzed three different LiDAR sensors currently used in the CAPTN Fjord5G project: a) Velodyne (VLP series), b) Blickfeld Cube 1, and c) Oyster OS2\footnote{Datasheets: \textbf{Velodyne} \url{https://velodynelidar.com/wp-content/uploads/2019/12/63-9229_Rev-K_Puck-_Datasheet_Web.pdf}, \textbf{Blickfeld} \url{https://www.blickfeld.com/wp-content/uploads/2022/10/blickfeld_Datasheet_Cube1-Outdoor_v1.1.pdf}, \textbf{Oyster} \url{https://data.ouster.io/downloads/datasheets/datasheet-revd-v2p0-os2.pdf}, last accessed: \today}. The goal was to understand how the raw data is typically captured and later processed~\cite{lidar_acq}.

The acquisition and transmission of raw LiDAR data often involves ROS (Robot Operating System) drivers that enable seamless integration and communication with LiDAR sensors. The data is then packaged into a special PointCloud/PointCloud2 ROS message, which can be serialized and sent over the ROS system or network, and then automatically viewed at the receiving point. All three vendors offer their UIs as desktop applications or Web GUIs that encapsulate all processes from acquisition to viewing under one solution and eliminate the need to set up the ROS framework. The raw data is often segmented and packed into UDP datagrams and sent over the network to monitor sensor output in real-time.

\subsection{Encoding}\label{section:4.2_data_encoding}
In general, encoding is the process of converting raw data into a compressed format that can be efficiently stored, transmitted, and decoded for playback on various media platforms and devices. This compression is essential to reduce the size of the transmitted data while maintaining acceptable quality, making it practical to store and distribute media content. We examine the main aspects of video and point cloud encoding, discuss how point cloud encoding could be reduced to the case of video, and finally discuss all available video codecs that we tested during field experiments in the CAPTN Fjord5G project and the optimal encoding parameters we found that can be used with GstWebRTCApp via corresponding GStreamer plugins for a smoother and higher quality playback experience.

\subsubsection{Point clouds encoding and streaming}
Point cloud codecs can be divided into two main groups~\cite{lidar_enc}:
\begin{itemize}
    \item [\textcolor{blue}{\textbullet}] Geometry-based. These codecs focus on partitioning the space of the raw point cloud into some compressed tree-based geometric structures such as OCTree (used by the PCL library~\cite{pcl}) or KD-Trees (used by the Draco codec from Google~\cite{draco}). The compression level is controlled by quantization parameters, which control the number of subdivisions of the trees from the root to each leaf node, resulting in a regular downsampling of the input clouds. 
    \item [\textcolor{blue}{\textbullet}] Video-based. Video-based codecs use techniques from traditional video coding to project the point cloud onto a set of planes, and then encode the projections, which include texture, depth, and an occupancy map, into the 2D domain~\cite{vpcc}. These projections are then treated as normal video frames and thus encoded in a typical way with H264/265 codecs (see details in the following subsection).
\end{itemize}

In addition to transformation capabilities, point cloud encoders perform another important task: they provide a way to serialize and segment point clouds. Unlike video frames, raw LiDAR data does not have built-in natural segmentation, and therefore the frequency cannot be easily controlled via the frames per second (FPS) parameter. Another disadvantage is that the encoder has to wait until the whole scene is captured before it can apply a geometry-based or video-based technique. 

Another important point is that these two techniques do not allow the live LiDAR data to be streamed as media over the WebRTC protocol. Although the video-based encoding technique theoretically allows this, in practice the output of the codec has multiple concurrent video streams that need to be transmitted and, more importantly, synchronized. This requires the development and integration of a dedicated decoder into the streaming pipeline. Since the AhoyRTC Director uses the decoding capabilities of browsers, a new point cloud decoder must be understood by browsers. This is an unrealistic requirement, so a custom non-browser WebRTC client must be developed in this case.

Summarizing all the information, we provide three possible ways to stream the live point clouds:
\begin{itemize}
    \item [\textcolor{green}{\checkmark}] \textbf{Uncompressed raw data streaming via integrated vendor solutions}. All LiDARs used in the CAPTN Fjord5G project have the instruments that provide the full pipeline from grabbing to viewing. While the advantages of this approach are clear because the dedicated vendor's solution is used, the disadvantages are high bandwidth requirements due to the uncompressed data, high latency even considering the absence of encoding due to the higher data rates and lack of WebRTC capabilities, and an isolated streaming solution for each sensor. 
    \item [\textcolor{green}{\checkmark}] \textbf{Non-media encoded data streaming over WebRTC data channels}. This way one of the encoding techniques can be used to segment and send the non-media data over WebRTC data channels. There are examples in the literature showing the successful applications of this way in real-time communication~\cite{lidar_dc}. Compared to the previous approach, it enables WebRTC capabilities and could be integrated under a single WebRTC platform. The disadvantage is that it is a non-media data transfer and to view the received packets, one still needs to implement a decoder and a viewer, which adds additional playout latency for unpacking and processing the received data.
    \item [\textcolor{green}{\checkmark}] \textbf{Rendering point clouds at the device, capturing and delivering as video over WebRTC}. This approach involves rendering LiDAR data on the computers installed on a ferry, capturing the rendered output, and delivering the resulting video streams over WebRTC. It reduces the amount of data transmitted, utilizes onboard computing resources, and allows the LiDAR output to be integrated into the AhoyRTC director. The disadvantage is the loss of granularity and detail in the transmitted data. The original three-dimensional spatial information can be simplified or lost due to the limitations of double encoding: the first for rendering on the device, the second for the resulting captured frames to be transmitted over WebRTC. As of the time of writing (March/April 2024), this approach is planned as the main one for streaming point clouds in real-time.
\end{itemize}

\subsubsection{Video codecs}\label{section:4.2.2_video_codec}
Video codecs (\textbf{co}der-\textbf{dec}oder pair) play a critical role in the compression, transmission, and playback of digital video content across various devices and platforms. It is a software or hardware algorithm, the latter is often called hardware acceleration (HA) and is done by modern GPUs. These codecs employ sophisticated compression techniques to reduce the size of video files while preserving perceptual quality, enabling high-quality video playback even over bandwidth-constrained networks. 

The existing codecs can be divided into three major groups:
\begin{enumerate}
    \item H group. The H group includes standards such as \textbf{H.264} (Advanced Video Coding, AVC), \textbf{H.265} (High-Efficiency Video Coding, HEVC). H.264, introduced in 2003, is nowadays the most widely used codec across various applications~\cite{codecs_pop}. H.265, a successor to H.264 released in 2013, offers improved compression efficiency, allowing for higher-quality video at lower bit rates.
    \item VP group. The VP group includes open-source video compression formats developed by Google. Notable members include \textbf{VP8} and \textbf{VP9}. VP8, released in 2008, is primarily used for web-based video streaming. VP9, introduced in 2013, offers improved compression efficiency over VP8, allowing for higher-quality video at lower bit rates. Both VP8 and VP9 are royalty-free and widely supported across platforms.
    \item AV1 group. The AV1 group is the next-generation video compression format developed by the Alliance for Open Media (AOMedia). Released in 2018, the \textbf{AV1} codec aims to provide even higher compression efficiency than existing standards such as H.265 and VP9, also remaining royalty-free and open source.
\end{enumerate}

\subsubsection{Performance evaluation of video codecs for real-time communication}\label{section:4.2.3_encoder_test}
To understand which codec is best suited to transmit a video stream in real-time via WebRTC from the "Wavelab" ferry to the AhoyRTC Director on shore, we conducted a series of experiments using our GstWebRTCApp. With the help of GStreamer plugins, we were able to test different encoding solutions during the whole day "Wavelab" trip across the Kiel Bay on 8. February 2024. For each solution, we turned on average (2 Mbps) and high resolution (6 Mbps) streams while the ferry was moving at high speed. We also thoroughly analyzed the technical aspects of video codecs to understand the limitations of their application to the CAPTN Fjord5G project setup.

In addition, we conducted a small independent objective test comparing different video codecs\footnote{Available at \url{https://github.com/gehirndienst/video-codecs-test}, last accessed: \today}. We took a publicly available video file and created GStreamer pipelines to encode this video data with different codecs into a Matroska container (.mkv file). To simulate real-time communication scenarios, we selected options to minimize latency and configured real-time presets for the encoders. The fixed 2Mbps bitrate for the encoded files ensures consistency across all codecs tested. The encoded videos are compared to the original video using Video Multi-Method Assessment Fusion (VMAF), Peak Signal to Noise Ratio (PSNR), and Structural Similarity Index (SSIM) metrics~\cite{codecs}. This test consists of two main components: GStreamer pipelines measure encoding time and file size, while FFmpeg commands compute the quality metrics~\cite{ffmpeg}. The tested codecs include H.264, H.265, VP8, VP9, AV1, H.264 hardware-accelerated (NVENC), and H.265 hardware-accelerated (NVENC).

\begin{figure}[!t]
\centering
    \includegraphics[width=\textwidth, height=6.6cm]{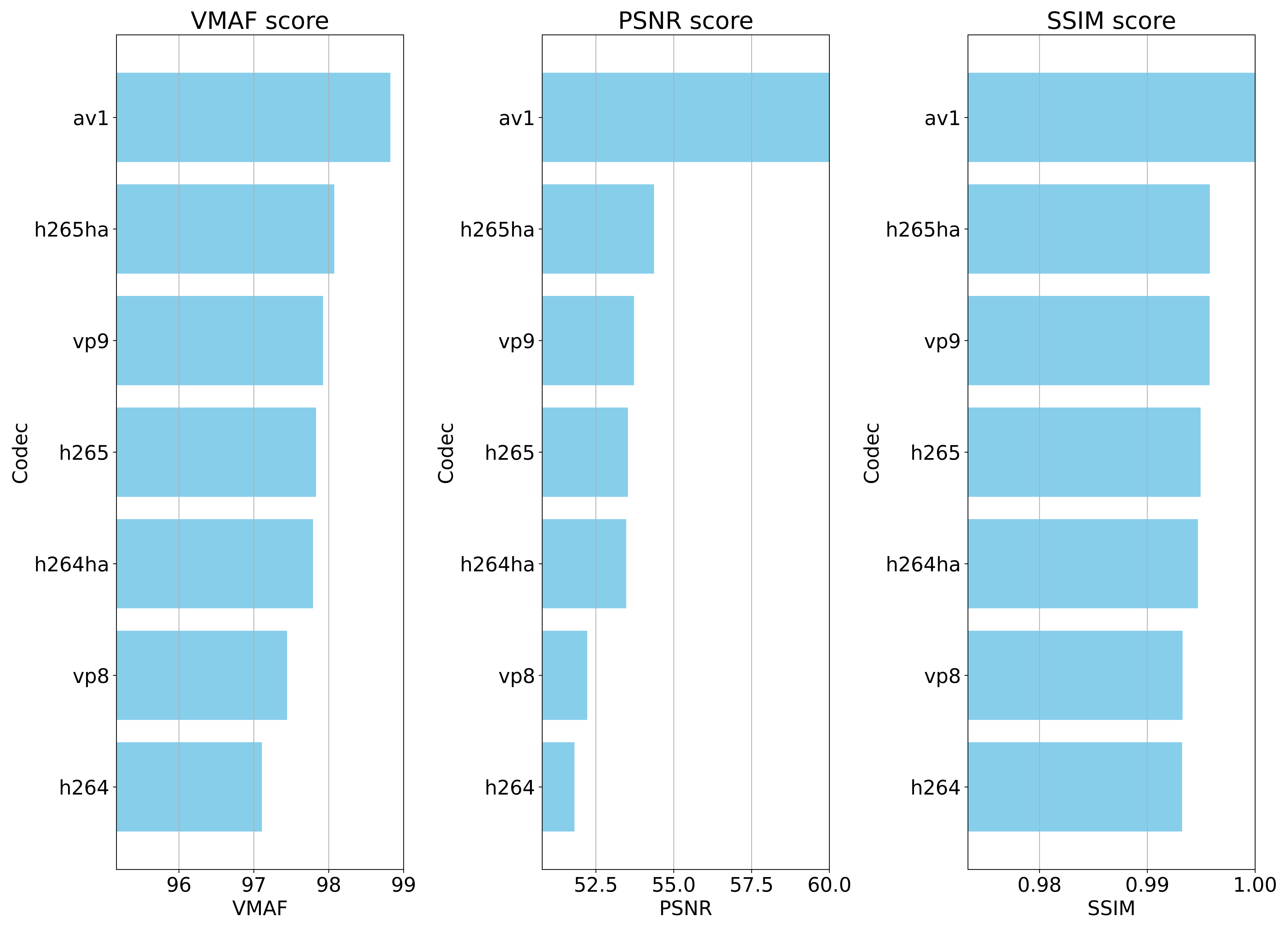}
    \caption{Video quality metrics for different video codecs on the same machine and with the same target bitrate and RTC tuning.}
    \label{figure:encoding_score}
\end{figure}

The results of the first evaluation are presented in Tab.~\ref{table:codec_comparison1}. We have grouped several aspects that represent the key features that need to be maintained at some point for consecutive real-time video transmission and analyzed each of them for each of the codecs. The results of the second evaluation are presented in Tab.~\ref{table:codec_comparison2} with encoding times and Fig.~\ref{figure:encoding_score} with video quality metrics for each of the codecs.

\begin{table*}[!ht]
\centering
\caption{Evaluation of video codecs based on "Wavelab" runs and technical analysis.}
\resizebox{\textwidth}{!}{\begin{tabular}{|c|c|c|c|c|c|}
\hline
\textbf{Feature} & \textbf{H.264} & \textbf{H.265} & \textbf{VP8} & \textbf{VP9} & \textbf{AV1} \\ 
\hline
Real-time stability & high & high & high & low & low \\
\hline
Stalling (video freezes) & sometimes & rare & sometimes & very often & very often \\
\hline
Bandwidth consumption & high & medium & high & medium & medium \\
\hline
Encoding speed & very high & high & high & low & very low \\
\hline
Viewable in AhoyRTC & yes & no & yes & yes & no \\
\hline
HA encoding support\footnotemark & yes & yes & no & no & potentially \\
\hline
Maximum resolution support & 4k DCI & 8k UHD & FHD & 8k UHD & 8k UHD \\
\hline
RTC fine-tuning capabilities & rich & rich & poor & poor & moderate \\
\hline
\end{tabular}}
\label{table:codec_comparison1}
\end{table*}
\footnotetext{Wavelab's server has only one GPU that can perform the HA for the video streams: NVIDIA 3090. It only supports H.264 and H.265 encoding/decoding acceleration. Newer NVIDIA 4000* GPUs also support AV1 encoding/decoding.}

\begin{table*}[!t]
\centering
\caption{Encoding time for 1 minute HD video for different codecs. "HA" in the table means H.-codecs with NVIDIA hardware acceleration.}
\resizebox{\textwidth}{!}{\begin{tabular}{|>{\centering\arraybackslash}p{2cm}|*{7}{>{\centering\arraybackslash}p{1.5cm}|}}
\hline
\textbf{Feature} & \textbf{H.264} & \textbf{H.265} & \textbf{VP8} & \textbf{VP9} & \textbf{AV1} & \textbf{H.264 HA} & \textbf{H.265 HA} \\ 
\hline
Encoding time, sec. & $6.80$ & $39.51$ & $16.27$ & $360.53$ & $201.90$ & $2.72$ & $3.81$ \\
\hline
\end{tabular}}
\label{table:codec_comparison2}
\end{table*}

Looking at the results, the following conclusions could be drawn:
\begin{itemize}
    \item [\textcolor{green}{\checkmark}] Across both evaluations, \textbf{H.264 and especially H.264 HA showed to be the most appropriate solution for the RTC considering the requirements of the CAPTN Fjord5G project}.
    \item [\textcolor{green}{\checkmark}] \textbf{H.265} and especially its accelerated version could be the best solution if it does not have problems with displaying in the AhoyRTC Director. It delivers better quality than H.264 (HA-version is top-2 across all metrics, see. Fig.~\ref{figure:encoding_score}) at a lower bitrate and supports 8k resolution. H.265 integration in modern browsers is still an ongoing process~\cite{h265_browsers}.
    \item [\textcolor{green}{\checkmark}] \textbf{VP8} could also be a good alternative to H.264, but the VP family was originally developed by Google to be used for watching videos on YouTube, there is no encoding HA support for these codecs, only decoding. VP8 takes much longer to encode but has better quality than software H.264 and worse than H.264 HA. \textbf{VP9} delivers the third best quality, but takes an enormous amount of time to encode, which was proven several times during the tests on the ferry when the streams suffered from extensive stalling. Therefore, the VP family is not suitable for the RTC in the CAPTN Fjord 5G project.
    \item [\textcolor{green}{\checkmark}] \textbf{AV1} has proven to be the most controversial option. While it delivers the best video quality at a bitrate comparable to H.265, could potentially be accelerated with the latest NVIDIA GPUs, and has RTC tuning capabilities, it currently does not work in AhoyRTC Director, has several issues with its GStreamer plugins, where AV1 encoder and payloader were developed by different teams and even in different languages, causing synchronization problems, and overall proved to be the most unstable solution requiring too much time for encoding. This can be explained by the fact that this codec is the newest one and is still in the active development and integration phase. At the time of writing (March-April 2024), it is not ready for use in the project.
\end{itemize}

\begin{figure}[!b]
\centering
    \includegraphics[width=\textwidth]{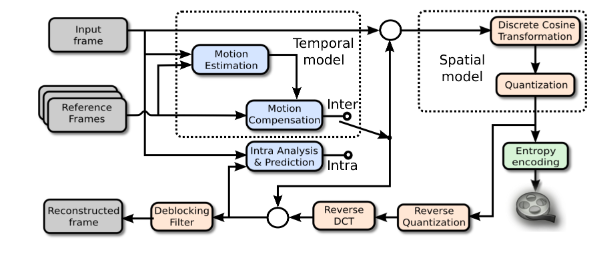}
    \caption{The H.264 codec operating scheme. Taken from~\cite{h264}.}
    \label{figure:h264}
\end{figure}

Considering the above conclusions, H.264 in its HA version was primarily selected as the main codec for video encoding in the CAPTN Fjord 5G project. Fig.~\ref{figure:h264} shows the workflow of the internal encoding and decoding processes. The H.264 codec combines several advantages, which are summarized below:
\begin{itemize}
    \item [\textcolor{green}{$+$}] \textbf{Popularity and compatibility}. H.264 remains the most widely used encoding solution today, supported by a wide range of devices, platforms, and software applications including web browsers.
    \item [\textcolor{green}{$+$}] \textbf{Real-time stability}. H.264 rarely experiences problems in real-time scenarios, making it one of the most stable solutions for applications that require continuous real-time video encoding and decoding.
    \item [\textcolor{green}{$+$}] \textbf{Versatility}. H.264 supports a variety of resolutions, frame rates, and profiles, making it suitable for a wide range of video applications, from high-definition to low-bandwidth real-time video streaming.
    \item [\textcolor{green}{$+$}] \textbf{Broad hardware acceleration support}. Most hardware platforms, including GPUs, CPUs, and dedicated chips, offer hardware acceleration for H.264 encoding and decoding. GStreamer provides stable plugins to enable HA for H.264.
    \item [\textcolor{green}{$+$}] \textbf{Vast tuning capabilities for real-time communications}. H.264 supports configurable internal optimizations to reduce latency and increase encoding speed. 
\end{itemize}

However, it also has some drawbacks that stimulated the development of modern codecs:
\begin{itemize}
    \item [\textcolor{red}{$-$}] \textbf{Limited efficiency}. While H.264 offers good compression efficiency, it can not match the compression ratios achieved by newer standards such as H.265, VP9, and AV1. As a result, H.264 may require higher bitrates to achieve a similar quality of experience, which can impact bandwidth consumption and increase latency.
    \item [\textcolor{red}{$-$}] \textbf{Limited maximum resolution}. H.264 can not process frames higher than 4K DCI ($4096\times2160$)~\cite{h264_4k}.
    \item [\textcolor{red}{$-$}] \textbf{High computational complexity}. H.264 processing can be computationally intensive, especially for high-definition and high-frame-rate video content. While hardware acceleration can mitigate this issue to some extent, it can still require significant processing time and resources.
\end{itemize}

\subsubsection{Video encoding optimization}
To optimize the performance of the H.264 HA video codec, we researched and experimented to find the best encoding parameters for the RTC. Since GStreamer is the main solution for adaptive communication, we used its \textbf{nvh264enc} plugin as the target encoding element. It has a list of properties (parameters) that can be configured depending on the particular scenario.

Based on our findings, we tweaked the following options, presented as key-value pairs, for our GStreamer H.264 HA encoder element (all others remain default and can be studied on the GStreamer website~\cite{gstreamer}):
\begin{itemize}
    \item [\textcolor{green}{\checkmark}] \textbf{$\text{preset=low-latency-hq}$}. The presets generally control the encoding techniques and the final image quality. The Low Latency High Quality preset is the only one NVIDIA recommends for video streaming~\cite{nvenc_nvidia}. It aims to minimize latency while maintaining high quality and includes options such as disabling bi-directional frames (B-frames) that reference both past and future frames, preferring inter-encoding over intra-encoding, which attempts to compress frames based on predicted frames for more efficient and faster compression. It also enables a single-frame VBV 2 pass. VBV (Video Buffering Verifier) is a mechanism used to control the bit rate of encoded video streams. This setting configures the VBV to operate in single-frame mode during the second pass of encoding.
    \item [\textcolor{green}{\checkmark}] \textbf{$\text{rc-mode=cbr}$}. The rate control mode tells the encoder how to compress frames to achieve the desired bitrate. Constant bit rate (CBR) is perfect for streaming applications where maintaining a consistent rate is critical. The other option, variable bit rate (VBR), allows the encoder to dynamically adjust compression based on the complexity of the scene, which is generally not needed for real-time streaming.
    \item [\textcolor{green}{\checkmark}] \textbf{$\text{gop-size=-1}$}. The Group of Pictures (GOP) size is the distance between two intra-coded frames (I-frames). These frames are independent of other frames and are sufficient by themselves to render a complete picture. Infinite (negative) GOP is best for streaming because it reduces the number of large I-frames that are transmitted, and at a higher bitrate, it is possible to improve quality without a linear increase in total bandwidth. A negative consequence is that if packets are lost, it may take longer for the decoder to recover.
    \item [\textcolor{green}{\checkmark}] \textbf{$\text{qos=true}$}. Setting this option to true instructs the encoder to handle Quality of Service (QOS) events from the rest of the pipeline. In GStreamer, QoS refers to the measurement and tuning of a pipeline's real-time performance. Handling these events improves the stability of pipeline elements.
    \item [\textcolor{green}{\checkmark}] \textbf{$\text{zerolatency=true}$}. Setting this option to true enables zerolatency tuning, which aims to optimize for fast encoding and low-latency streaming. This tuning enables multi-slice encoding, so that a frame contains multiple encoding units, while also reducing look-ahead, the option to control how many future frames are "looked at" in addition to the current frame during the compression process.
\end{itemize}

\subsection{RTP payloading}
In the context of multimedia streaming with WebRTC using GStreamer, the RTP payloader performs the critical task of encapsulating media data into RTP packets. Encoded video frames are broken down into smaller, more manageable units and packaged into RTP packets with the appropriate header for transmission over the network. The payloader ensures that the media content is formatted according to the specifications and requirements of the RTP protocol, which is essential for interoperability and compatibility with WebRTC endpoints~\cite{rtp_payload}. It also aims at optimizing the latency and facilitates synchronized playback at the receiving end.

Given our decision to use the H.264 encoder, we turn to GStreamer's RTP payloader, designed specifically for this codec: \textbf{rtp264pay}. As an encoder, it offers some adjustable properties to fine-tune its behavior. During our experiments, we found that tweaking the following helped to improve real-time performance:
\begin{itemize}
    \item [\textcolor{green}{\checkmark}] \textbf{$\text{auto-header-extension=true}$}. This is a technical property that enables automatic header extension handling, allowing additional header information to be included in RTP packets. For example, to use GCC congestion control (see Sec.~\ref{section:3.2_webrtc_protocol}), one must enable the transport-cc RTP extension to add a transport-wide sequence number to the RTP header needed to receive TWCC packets~\cite{twcc}.
    \item [\textcolor{green}{\checkmark}] \textbf{$\text{mtu=1250}$}. It defines the Maximum Transmission Unit (MTU), i.e. the maximum size for the RTP packets. By setting the MTU size to 1250 bytes, we aim to optimize packetization to fit within the MTU limit of the VPN used on the Wavelab ferry, thus reducing the likelihood of packet loss or fragmentation.
    \item [\textcolor{green}{\checkmark}] \textbf{$\text{config-interval=1}$}. This property specifies the interval in number of frames at which configuration information, such as SPS (Sequence Parameter Set) and PPS (Picture Parameter Set) for H.264 codec, is transmitted within the RTP stream. When set to 1, configuration data is sent with every I-frame, ensuring that decoders have up-to-date codec configuration for proper decoding.
    \item [\textcolor{green}{\checkmark}] \textbf{$\text{aggregate-mode=zero-latency}$}. As with nvh264enc, this option enables zerolatency tuning. In this mode, the payloader uses aggressive packetization strategies to encapsulate individual video frames into RTP packets without introducing significant buffering or delay. Rather than waiting for a certain number of frames to accumulate before packetizing, as in traditional schemes, the zerolatency-tuned payloader dynamically adjusts the packetization process to minimize packetization overhead while ensuring that each packet contains enough video data for decoding at the receiver end.
\end{itemize}

\subsection{Latency}
Latency refers to the time it takes for data to travel through a communication medium from its source to the final destination. It plays a critical role in real-time communications, where even small delays can significantly influence the user experience, as high latency can result in noticeable delays between actions and their effects, leading to communication breakdowns, reduced engagement, and a diminished sense of presence. Minimizing latency is a key objective in the design and operation of real-time communication systems.

Final latency consists of four parts: a) \textbf{processing latency}, which is the time it takes to acquire, encode, and pack data; b) \textbf{queuing latency}, which is the delay that occurs when data packets are queued before being processed or transmitted, c) \textbf{transmission latency}, which is the time it takes to transmit a packet over a communication medium and d) decoding latency, which is the time it takes to decode the incoming packets. We examined all of these parts with our approach presented in this section and also experimented with measuring the transmission latency from "Wavelab" to the shore-based AhoyRTC Director to understand if our data transmission system meets the requirements for seamless RTC.

\subsubsection{Processing and queuing latency}
We have combined these two parts into one subsection because they are fully covered by our "GstWebRTCApp" application. Processing and queuing latencies include all delays caused by the actual processing of the video stream through the GStreamer pipeline. This includes acquisition, encoding, and payload, which are covered in the previous subsections. 

The acquisition is done by some source elements, the default source being an RTSP stream as discussed in Sec.~\ref{section:4.1.1_video_acq}. The corresponding GStreamer element is \textbf{rtspsrc}. It takes a URL of the source data and handles its delivery in the same way as the WebRTC receiver: it processes the data as an RTP stream. Every WebRTC sender and receiver has a jitter buffer. It is a mechanism to smooth out variations in packet arrival time by temporarily storing incoming packets and releasing them at a steady rate to the playback system~\cite{jitter_buffer}. 

The rtspsrc element has a latency property that controls the size of a jitter buffer in milliseconds. The default is 2 seconds. This is too much for RTC and is not needed in the CAPTN Fjord5G project because all processing is done on the CAU server installed on the "Wavelab" where all cameras are located and all participants are within the same ship network. However, setting the latency to zero or very low values results in a "wobbling" video. We have experimented with this value and found that \textbf{100 milliseconds} is a good compromise value to eliminate the negative effects and properly process the camera streams without adding too much unnecessary delay. Moreover, if there is no current problem with RTSP acquisition, GStreamer will automatically redistribute the jitter buffer latency to lower values.

Encoding with our chosen H.264 HA encoder is very fast and adds \textbf{less than 1 millisecond} as shown in Sec.~\ref{section:4.2.3_encoder_test}. The payloader itself does not add any significant latency either. However, the last and most important element that manages the WebRTC part of the GStreamer pipeline, \textbf{webrtcbin}, also has a latency property that controls its jitter buffer and is set to 200 milliseconds by default. However, since we do not need to wait for a certain number of packets to be ready to send, while all packets should be sent immediately, this latency could be minimized. We set it to \textbf{10 milliseconds} after some experimentation.

\subsubsection{Transmission latency}
Transmission latency is the most uncontrolled part of the latency chain. It depends on multiple factors such as signal quality, traffic overhead, congestion, propagation, transport mechanisms, etc., which are combined under a single definition of "network conditions".

Another important issue is that WebRTC leaves no way to calculate one-way delay between the sender and receiver because all incoming RTP packets only contain the RTP timestamp, which is randomly initialized at the beginning of each session and then incremented by a certain amount for each video frame, not packet~\cite{rtcp}. This means that two consecutive packets of the same video frame will have the same RTP timestamp, making it impossible to know the absolute transmission time for each packet. The only way to synchronize clocks is to analyze RTCP reports because they contain an additional Network Time Protocol (NTP) timestamp that represents the wall clock time of the sender or receiver. RTCP reports provide Round Trip Time (RTT) values for each WebRTC stream. Note that these RTT measurements are affected by the stream bitrate.

To overcome this, we implemented a utility in Golang that measures the RTT between the sender and the AhoyRTC Director's TURN server by sending periodic probes and sharing the value via the MQTT topic\footnote{Available at \url{https://github.com/gehirndienst/gstwebrtcapp/tree/main/tools/rtt-checker}, last accessed: \today}. By measuring it from the "Wavelab", we got average RTTs between 50 and 80 milliseconds. The one-way delay could be roughly assumed to be half of the RTT~\cite{rtt_delay}. So the transmission latency \textbf{on itself} is about \textbf{30-40 milliseconds}. As mentioned above, this value will be higher for high bitrate streams due to the additional network queuing.

\subsubsection{Decoding latency}
The decoding latency consists of two parts. First, the latency is caused by the jitter buffer at the receiver side, and second, the processing latency is caused by the decoding process itself. For the latter, decoding is generally faster than encoding, so the latency can also be expected to be \textbf{less than 1 millisecond} even without hardware acceleration~\cite{enc_dec}.

As for the jitter buffer on the receiver side (browser), it is hard to say what the latency is there because it is not controlled by the user and is a black box. Looking at the source code of Google Chrome, it could be noticed that the initial value is \textbf{30 milliseconds} and then it can grow based on jitter estimates depending on network conditions\footnote{Google Chrome WebRTC Jitter Estimator \url{https://webrtc.googlesource.com/src/webrtc/+/f54860e9ef0b68e182a01edc994626d21961bc4b/modules/video_coding/jitter_estimator.cc}, last accessed: \today}.

\subsection{Summary}
This section discusses the entire multimedia data processing scheme: from the acquisition of raw data from sensors to the display of the results on the receiver's browser. We have described the key aspects of processing video and point clouds using our adaptive streaming application based on GStreamer.

We started with the data acquisition phase, highlighting how it happens for video and LiDAR streams. We discussed the general issues related to the transmission of point clouds and summarized three possible ways to deliver them: a direct transmission over UDP, and two variants of WebRTC-based transmission over data channels and as a video capture rendered on the ship's server. 

We then analyzed the key aspects of video and point cloud encoding. We presented all state-of-the-art methods and performed performance tests for the video codecs, resulting in the selection of the hardware-accelerated version of the H.264 codec as the best. Based on the results of the experiments, certain properties of the encoding and payloding elements were optimized to minimize processing latency. 

The latency itself was discussed in detail in the last subsection with the experimental measurements of the latency between the sender and the AhoyRTC director using our self-developed probing service. We also discussed the tuning of the latency parameters for capturing and sending the video stream.

Summarized conclusions:
\begin{itemize}
    \item [\textcolor{green}{\checkmark}] Both video and LiDAR streams can be captured in a format that allows further processing (video frames and point clouds), while they can also be captured as raw and processed data.
    \item [\textcolor{green}{\checkmark}] 3 ways to stream the point clouds (vendor solution and 2 variants of WebRTC streaming) have been studied, where \textbf{streaming rendered point clouds as video over WebRTC} was chosen as the best because it allows to view them in AhoyRTC Director and manage them as a normal video stream with "GstWebRTCApp".
    \item [\textcolor{green}{\checkmark}] According to the results of our performance tests of 5 different video codecs and two of them with hardware acceleration, the \textbf{H.264 hardware accelerated video codec was selected as the primary one} to be used for all video streams.
    \item [\textcolor{green}{\checkmark}] After much experimentation with sending the video streams from "Wavelab" to the AhoyRTC Director via WebRTC, \textbf{the encoding and payload elements were optimized to minimize the resulting latency} with our description and explanation of all the optimized properties.
    \item [\textcolor{green}{\checkmark}] \textbf{The latency between the Wavelab and the AhoyRTC Director} has been measured and reported with our probing tool and \textbf{is typically around 30-40 milliseconds under stable network conditions}.
    \item [\textcolor{green}{\checkmark}] The latency for all processing steps has been calculated and defined. The outcome leads to the conclusion that \textbf{the resulting latency should not exceed 300 milliseconds under the typical network conditions observed in the Bay of Kiel}. Therefore, the stalling effect should be very rare and the quality of experience for real-time communication should remain at a high level.
    \item [\textcolor{green}{\checkmark}] In a congested network, latency can reach very high values, so an adaptive transmission mechanism is required. This is discussed in the next chapter.
\end{itemize}

\clearpage
\section{AI-enhanced adaptive data transmission}
One of the key challenges for real-time media delivery over the WebRTC protocol in 5G networks is the variability of network conditions, including bandwidth availability, latency, and packet loss rates. These conditions can vary rapidly due to the stochastic nature of cellular networks. Adaptive communication techniques that dynamically adjust the quality of transmitted video streams, such as bit rate, based on real-time network feedback are essential to ensure optimal quality of experience (QoE) for end users. 

By using such mechanisms, live streams can be adapted to changing network conditions, maximizing the usage of available network resources while maintaining an acceptable level of QoE. For example, during periods of high network congestion or limited bandwidth availability, adaptive communication algorithms can intelligently reduce the bitrate or resolution of media streams to prevent buffering and minimize latency, ensuring smooth playback and real-time interaction. 

In this section, we examine the baseline solution for congestion control in the WebRTC protocol and present our AI-based solution for adaptive data transmission.

\subsection{Congestion control}
Congestion occurs when the demand for network resources exceeds their capacity due to limited bandwidth at some bottlenecks along the network paths, resulting in delays, packet loss, and degraded quality of service. Congestion control mechanisms mitigate these problems by dynamically adjusting data transmission rates and optimizing resource utilization. These mechanisms often involve feedback-based algorithms that monitor network conditions, detect congestion events, and adjust transmission parameters accordingly to alleviate congestion.

There is an RTP Media Congestion Avoidance Techniques (RMCAT) working group that aims to develop new protocols to manage network congestion in the context of RTP streaming~\cite{rmcat}. As a result, several algorithms have been proposed by different vendors. Among them are:
\begin{enumerate}
    \item Network-Assisted Dynamic Adaptation (NADA) by Cisco.
    \item Self-Clocked Rate Adaptation for Multimedia (SCReAM) by Ericsson.
    \item Google Congestion Control Algorithm for Real-Time Communication (GCC) by Google.
\end{enumerate}

As mentioned in Sec.~\ref{section:3.2_webrtc_protocol}, the most widely used implementation of the WebRTC protocol implements the GCC algorithm~\cite{gcc}. The core principle of the algorithm is to use the queuing delay gradient to infer congestion. Queuing delay is calculated as the ratio of the length of the queue to the capacity of the bottleneck link, and GCC attempts to minimize the size of the queue without underutilizing the link. To accomplish this, the algorithm probes for available bandwidth by increasing its transfer rate until a positive change in queuing delay is detected. It uses two controllers to do this: a) a delay-based controller that calculates the rate for the video stream using a Kalman filter that takes into account the most recent throughput and delay measurements, and b) a loss-based controller that calculates the final rate by increasing or decreasing the proposed rate with the fixed coefficients depending on the current packet loss. By combining delay-based and loss-based control mechanisms, GCC ensures a robust and efficient data transmission rate~\cite{gcc2}.

The GCC algorithm has certain advantages and disadvantages. It responds well to packet loss, aiming to not break the transmission at any point, which is valuable in RTC. It has also proven to be fair between multiple streams~\cite{congestion_control}. However, its main disadvantage is a very slow re-stabilization after congestion. The stream rate increases slowly, leading to the underutilization of the network link. That's why we aimed to develop AI-based solutions to complement GCC's strengths and address its limitations, ensuring more efficient and adaptive congestion control in RTC scenarios.

\subsection{Deep reinforcement learning approach}
\begin{figure}[!b]
\centering
\includegraphics[width=1.0\columnwidth, height=5.0 cm]{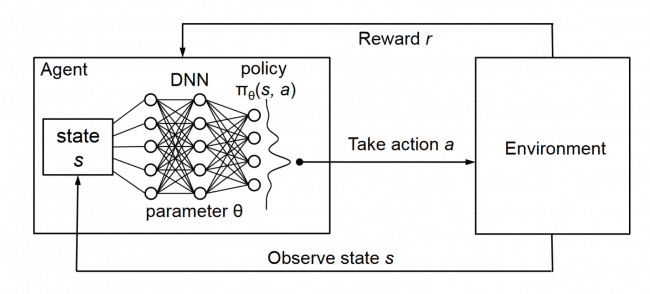}
\caption{Deep Reinforcement Learning workflow.} 
\label{figure:drl_schema}
\end{figure}
Reinforcement learning (RL) is a type of machine learning (ML) in which an agent is trained to learn through interaction with the environment and from its actions. The agent receives feedback through rewards or penalties for each action it takes. Markov Decision Process (MDP) is a mathematical framework used to model decision problems~\cite{sutton}. It assumes that the current state of the environment contains all relevant information needed to make a decision and that the environment follows the so-called Markov property, i.e. the process is memoryless: the probability of selecting the next state depends only on the current state. Deep Reinforcement Learning (DRL) is a further development of this strategy, using Deep Neural Networks (DNN) to approximate the agent's policy and value functions in MDP. The scheme of the whole process is shown in Fig.~\ref{figure:drl_schema}.

In DRL, the agent observes the current state of the environment, chooses an action based on its learned policy, receives a reward or penalty based on that action, and updates its policy accordingly through the training process. DRL can be applied to environments with both continuous and discrete action spaces. For adaptive transmission, agents can select actions from a predefined set of well-known quality presets, such as HD, FullHD, 4K, etc., with predefined bitrate, frame rate, and resolution, or they can select a floating point value(s) representing the target bitrate, providing fine-grained control over video streaming quality.

We chose model-free DRL as our main approach. Model-free means that there is no prior knowledge about the environment and no model of it. This approach has the following advantages
\begin{itemize}
    \item [\textcolor{green}{$+$}] It learns from its actions and does not require labeled data or expert knowledge as in supervised learning. Both are difficult to obtain for the problem of optimizing data transmission in RTC.
    \item [\textcolor{green}{$+$}] Proactive nature, allowing agents to take actions in anticipation of future states while working in assistance mode, providing the flexibility to defer actions to e.g. switch to manual control until the next state is available.
    \item [\textcolor{green}{$+$}] Adaptability to dynamic, real-world network conditions, which, especially for cellular networks, are highly stochastic and often vary in unpredictable ways.
\end{itemize}

The main drawback is that the outcome of the model depends on the manual fine-tuning of the agent's behavior, which is directly influenced by the design of the reward function.

\subsection{Performance evaluation of AI-enhanced adaptive communication}
As a DRL algorithm, we used the Soft Actor-Critic (SAC) model~\cite{sac}. It is a modern off-policy algorithm that emphasizes entropy and learns with a stochastic actor. A key feature of SAC, and a major difference from other common DRL algorithms, is that it is trained to maximize a trade-off between expected return and entropy, a measure of randomness in the policy. It works only in environments with continuous action spaces and incorporates a stochastic policy, allowing it to explore a wide range of actions and adapt effectively to varying network conditions. This adaptability is essential in a real-world RTC scenario where network conditions are highly dynamic.

As a state, we used several WebRTC metrics, including RTT, PLR, jitter, retransmission and goodput measurements. As an action we have a target bitrate value that is immediately applied to the stream. The maximum bitrate was set to $10$ Mbps while the minimum is $0.4$ Mbps.

As in Sec.~\ref{section:3.2.2_sandbox}, we used our sandbox scenario to model different network experiences in "Gymir5G". Here we have three types of scenarios: easy, moderate, and hard. The "easy" one represents a network without any significant problems for data transmission and rare packet loss. The "moderate" scenario repeats a congested scenario from Sec.~\ref{section:3.2.2_sandbox}: it experiences several moments of congestion. The "hard" scenario assumes network disruptions, extreme interference and noise, and constant movement in the non-line-of-sight (NLOS) direction, all of which combine to cause extreme delays and losses. Sometimes, not even a single packet could be delivered within a period until the next decision should be made. The model was trained only on "easy" and "moderate" scenarios so that the "hard" scenario also tests the ability to work in unseen extreme circumstances.

\begin{figure*}[!b]
\centering
\includegraphics[width=1.0\textwidth, height=5.7cm]{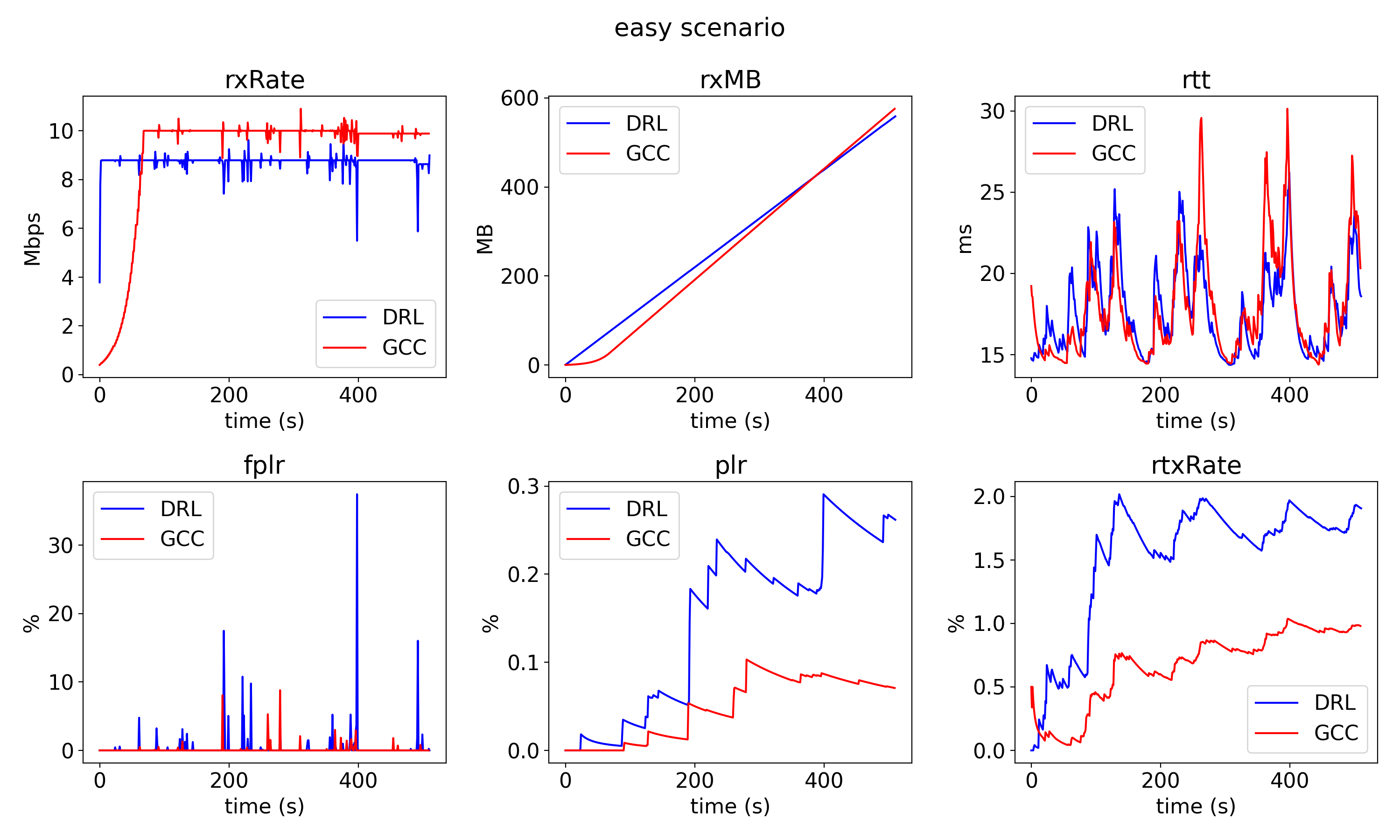}
\caption{Comparison plots of key features of DRL and GCC for the "easy" scenario.} 
\label{figure:easy}
\end{figure*}

\begin{figure*}[!b]
\centering
\includegraphics[width=1.0\textwidth, height=5.7 cm]{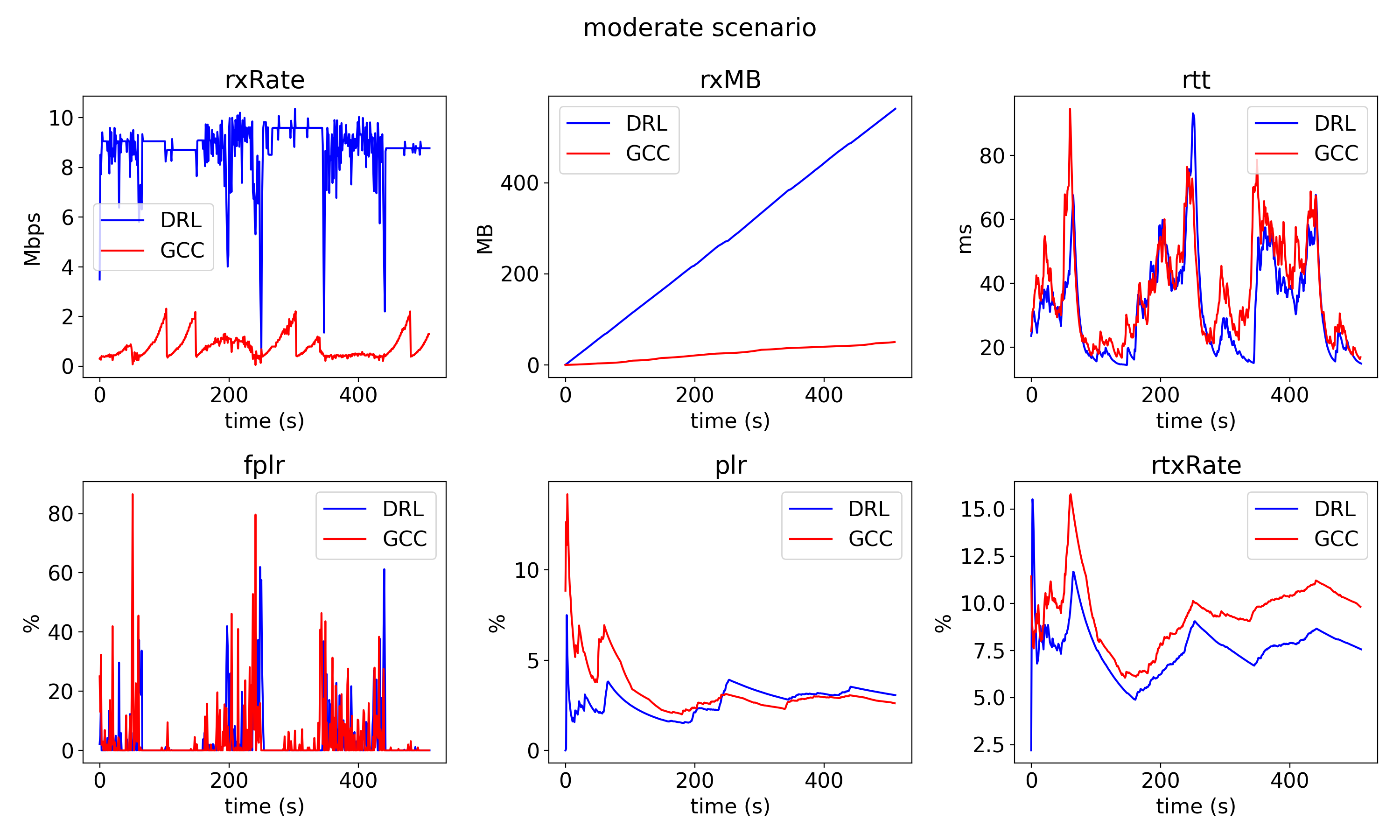}
\caption{Comparison plots of key features of DRL and GCC for the "moderate" scenario.} 
\label{figure:moderate}
\end{figure*}

\begin{figure*}[!t]
\centering
\includegraphics[width=1.0\textwidth, height=5.7 cm]{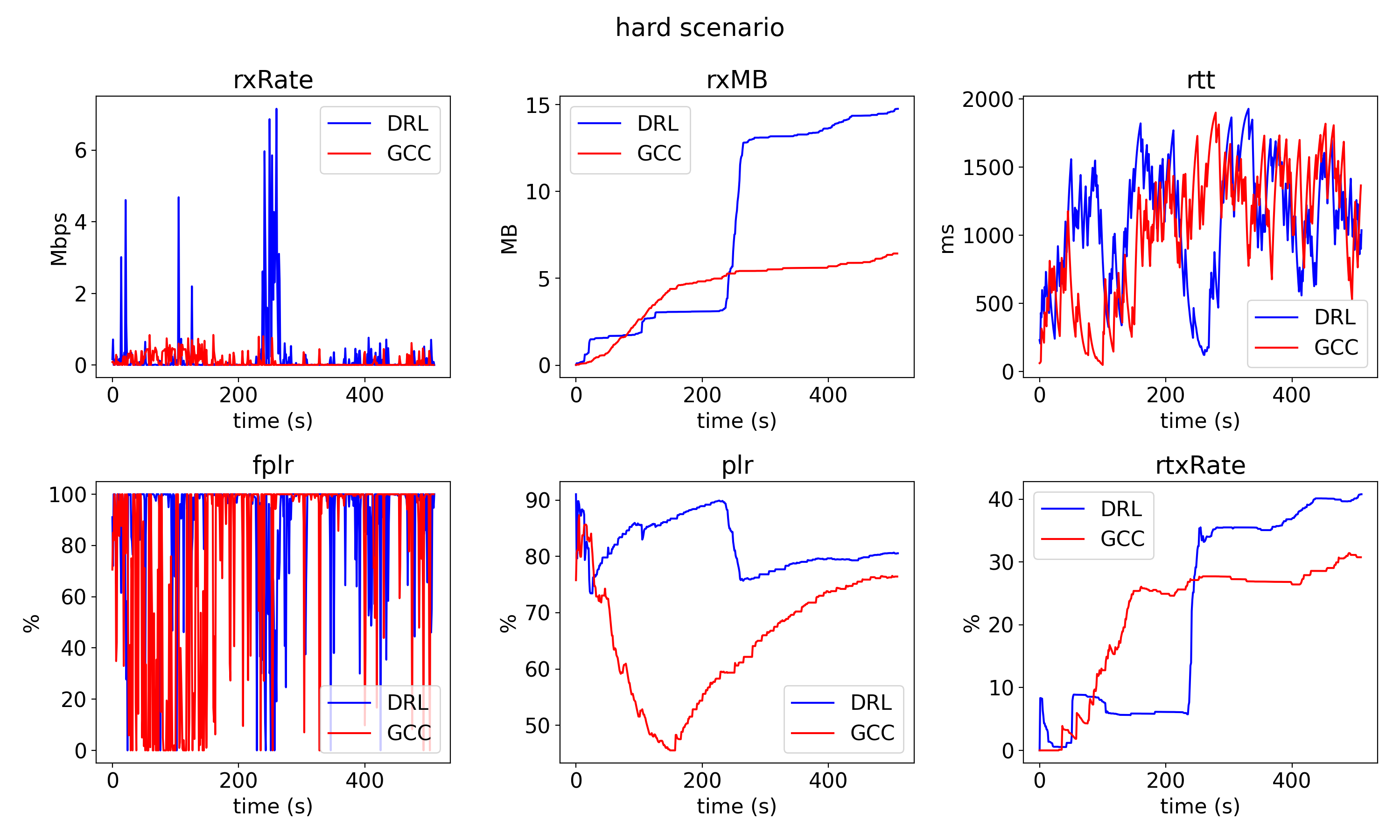}
\caption{Comparison plots of key features of DRL and GCC for the "hard" scenario.} 
\label{figure:hard}
\end{figure*}

We evaluated the performance of the DRL algorithm in comparison with the GCC~\cite{smirnov24}. The results are presented on three plots: Fig.~\ref{figure:easy}, Fig.~\ref{figure:moderate} and Fig.~\ref{figure:hard} for each scenario.

The six key features are plotted: the current receive rate in Mbps, the global amount of Mbytes received, the current round-trip-time in milliseconds, the current and global packet loss rate, and the global retransmission rate in \%.

Looking at the plots, the following conclusions could be made:
\begin{itemize}
    \item [\textcolor{green}{\checkmark}] "Easy" scenario: GCC is slightly better. It has a bit lower PLR (0.15\% difference), and it slowly converges to the maximum rate and stays there with slight drops. DRL is a bit conservative here (90\% of the rate) and compensates with a lower RTT, also almost indistinguishable in absolute values (5 milliseconds difference). This could be explained by the fact that SAC applies a special policy: not to change the target bitrate if a new action does not differ more than 10\% from the current bitrate to stabilize the real performance.
    \item [\textcolor{green}{\checkmark}] "Moderate" scenario: DRL vastly outperforms. This shows the conservative nature of GCC, which is analyzed in detail in~\cite{gcc2}. It cannot exceed rates higher than 3 Mbps because every time a new small congestion occurs, GCC immediately cuts off the rate. On the contrary, DRL does not hesitate to rapidly increase the rate at the cost of sharp drops of more than 50\% of the rate. GCC simply cannot do this because it is based on a trendline filter that increases the next estimated value very slowly. Despite a big gain in rate, DRL has almost the same local and global PLR and overperforms in RTT as well.
    \item [\textcolor{green}{\checkmark}] "Hard" scenario: DRL is better. GCC simply stays at the lowest possible rate with weak attempts to increase it, while DRL uses a different strategy that gives it an advantage in the "moderate" scenario. It tries to find windows to raise the rate high enough to provide decent QoE, at least for a short time. The main attempt was made between 200 and 300 seconds, and the scenario responds with an explosive loss that drops the rate to GCC levels. After this splash and a significant drop in reward, the DRL model does not attempt to cope with low rates and follows almost the same rate policy as GCC, with a stabilizing trend in PLR and a lower RTT than GCC.
\end{itemize}

\subsection{Summary}
This section discusses adaptive communication for media data transmission in RTC. We presented the traditional WebRTC congestion control algorithm GCC and our AI-enhanced approach. We compared the performance of both on the sandbox scenario and discussed the results obtained.

Summarized conclusions:
\begin{itemize}
    \item [\textcolor{green}{\checkmark}] Real-time media delivery over WebRTC in 5G networks requires adaptive control for better performance and reliable ship-to-shore communication to avoid network congestion that disrupts continuous data delivery.
    \item [\textcolor{green}{\checkmark}] The WebRTC protocol uses a standard congestion avoidance algorithm called GCC. "GstWebRTCApp" integrates GCC bandwidth estimates and can deliver them via MQTT. However, \textbf{GCC can be very conservative in terms of rate control}, resulting in underutilization of bandwidth.
    \item [\textcolor{green}{\checkmark}] The deep reinforcement learning approach allows learning the optimal data transmission from experience. We showed that the properly trained and tuned model \textbf{outperforms GCC}. This model is also integrated into "GstWebRTCApp" and is ready to use.
    \item [\textcolor{green}{\checkmark}] \textbf{The combination of congestion control and an AI-enhanced model can result in an optimal adaptive communication solution}.
\end{itemize}

\clearpage
\section{Conclusion}
This conceptual analysis covers all steps and layers of data transmission from a moving ship to a shore-based control center in a real-time communication scenario in the context of the scientific project CAPTN Fjord5G. Some experiments are specifically designed for the Bay of Kiel as the main location of the project. We presented simulation-based experiments as well as our GStreamer-based solution for real data transmission, which at the time of writing (March-April 2024) is already deployed on the university's server located on the ferry "Wavelab".

The sections presented in this research address the following questions and topics:
\begin{enumerate}
    \item Introduction, objectives, and software. It gives a sketchy overview of the project setup and the referencing ferry, our list of questions and goals for this conceptual analysis, while also presenting the solutions we developed in the scope of the project: a) "Gymir5G" - a 5G simulation with WebRTC features and AI-based assistance support, b) "GstWebRTCApp" - a GStreamer-based Python application for adaptive media delivery over WebRTC protocol. Both are publicly available and links can be found in the respective subsections. 
    \item The docking scenarios in the "Schwentine" area of the Bay of Kiel. The main scenario is ship-to-shore uplink data transmission from the moving ship to the shore-based control center, while we also evaluated downlink and multi-home (5G + Wi-Fi) scenarios. All of them were simulated in "Gymir5G". This section presents detailed estimates of the available bandwidth, latency times, and overall reliability of communication in the mentioned area, while also explaining the main network issues such as handover or congestion that may affect data transmission.
    \item The WebRTC protocol. We explained why it was chosen as the main protocol for the project, how it works, and why it is considered the best for real-time communication scenarios. We also discussed the difference between live and non-live data delivery. Finally, this section presents a series of WebRTC experiments performed in "Gymir5G" to evaluate the performance of the main WebRTC features and to demonstrate the effectiveness of WebRTC in real-time communication over the standard transport protocols.
    \item The entire data processing pipeline. This section completely covers the transmission of the media data: the video and LiDAR streams. We described the acquisition, encoding, packaging, and transmission phases. The goal is to show how the data flows from the sensor to be displayed in the AhoyRTC Director - a WebRTC client from Addix GmbH used in the CAPTN Fjord5G project. The acquisition and encoding of point clouds and video frames are discussed in detail. We evaluated the performance of different video codecs and found that the hardware-accelerated H.264 codec is the best. We also experimented with real ship-to-shore data transmission to present the optimized encoding, payload, and transmission parameters of the corresponding GStreamer elements used in our "GstWebRTCApp". Finally, we estimated the common latency at all stages when employing "GstWebRTCApp" for media data transmission.
    \item Adaptive real-time communication. The goal of this section is to present the traditional and AI-based approaches to adapting real-time data transmission to changing network conditions. We described our deep reinforcement learning-based solution, showing how exactly it controls the media stream, and performed a comparison with the standard WebRTC congestion control algorithm GCC. The results show that the AI model outperforms GCC. Both could be used together or separately in "GstWebRTCApp".
\end{enumerate}

\clearpage
\let\Section\section 
\def\section*#1{\Section{#1}}
\bibliographystyle{ieeetr}
\bibliography{references}

\end{document}